\documentclass[prd,showpacs,nofootinbib,preprint,10 pt]{revtex4}
\usepackage{amsmath,amsfonts,float,graphicx,color,xcolor,epsfig}
\usepackage{epstopdf}
\usepackage{hyperref}
\usepackage{slashed}

\begin{document}
\begin{titlepage}

\begin{flushright}
Alberta Thy 10-19\\
\end{flushright}

\begin{center}
\setlength {\baselineskip}{0.3in} 
{\bf\Large\boldmath 
Vertex renormalization and hard scattering symmetry breaking corrections to $B$ to axial vector meson form factors at large recoil
}\\[5mm]
\setlength {\baselineskip}{0.2in}
{\large  Arslan Sikandar$^1$, M. Jamil Aslam$^{1,2}$, Ishtiaq Ahmed$^{3}$, \\ 
        and Saba Shafaq$^4$}\\[5mm]
        
$^1$~{\it Physics Department, Quaid-i-Azam University,\\ Islamabad 45320, Pakistan}\\[5mm] 

$^2$~{\it Physics Department, University of Alberta,\\ Edmonton, T6G2E1, Alberta, Canada\footnote{Visiting Professor}}\\[5mm] 

$^3$~{\it National Centre for Physics, Quaid-i-Azam University Campus,\\
          Islamabad 45320, Pakistan}\\[5mm]
$^4$~{\it Department of Physics, International Islamic University (IIU),\\
 Islamabad, Pakistan.}\\[3cm] 

\end{center}

{\bf Abstract}\\[5mm] 
\setlength{\baselineskip}{0.2in} 
The symmetries arise due to heavy quark and large energy limit help us to reduce the number of independent form factors in the heavy-to-light $B$-meson decays. It is expected that these symmetry relations are not exact and are broken by the perturbative effects, namely, the vertex corrections and the hard-spectator scatterings. The former are included in the form factors via vertex renormalization whereas the later are calculated through light-cone distribution amplitudes. We first calculate these symmetry breaking corrections to the form factors involved in semileptonic $B$-meson to an axial-vector $(K_{1})$-meson decay. Later, by using these form factors we see their effect on the physical observables such as the zero-position of the forward-backward $(\mathcal{A}_{FB})$ asymmetry and the longitudinal lepton polarization ($P_L$) asymmetry in $B\rightarrow K_1(1270) \mu^+\mu^-$ decay. We find that as a result of these corrections to the form factors, the zero-position of the forward-backward asymmetry is shifted by $10\%$ from its SM value while the effects on $P_L$ are rather insignificant.

\end{titlepage}

\section{Introduction}\label{intro}
Form factors for $B$-meson decaying to a light-meson $f$, where $f$ can be a pseudoscalar $(P)$, a vector $(V)$, an axial-vector $(A)$ or a tensor $(T)$ meson, arise due to the matrix element of local flavor-changing currents (FCC) $\bar{q}\Gamma q$ with $\Gamma$ representing some spin-structure. The hadronic form factors play a crucial role in the accurate predictions of some physical observables (e.g.,  branching ratios, angular coefficients, etc.) in different semileptonic $B$-meson decays.  At present, several measurements of $B$ decays involving both flavor-changing neutral currents (FCNC) $ b\to (d, s)\ell^+\ell^-$ and flavor-changing-charged-current $(b\to c \ell \nu_\ell)$ have shown the possible hints of physics beyond the standard model (SM) (see e.g., \cite{Albrecht, Dyk} and references therein). Despite the accurately known form factors for $B\to K^*$ and  $B\to D^*$, the efforts to make them more precise are still focus of the ongoing theoretical and phenomenological studies \cite{Dyk2}.

In heavy-to-light decays, the form factors are mainly dominated by the QCD interactions at small momentum transfer and hence they can not be computed in the perturbation theory. Effective theories embedding certain symmetries are used to reduce the number of independent form factors. In this context, one of the symmetries in the decays of mesons containing heavy quarks is known as the heavy quark symmetry (HQS) which is based on an expansion in the inverse powers of the heavy quark
mass. By invoking this symmetry expansion, one can get certain symmetry relations \cite{Isgur:1989vq, Isgur:1989ed, Neubert:1993mb} which may not
be evident in the full QCD. These relations are used to relate the matrix elements corresponding to different currents and hence reduce the number
of independent form factors \cite{Charles:1998dr, Grozin:1996pq, Georgi:1990um}. For instance in $B \rightarrow D^*$
the seven independent form factors can be reduced to single Isgur Wise function \cite{Isgur:1989vq}.

In case of semileptonic decays $B\to (\pi, \rho, K^{*})\ell^{+}\ell^{-}$, when energy $(E)$ of the final state meson is large, one can make the expansion in the powers of $1/E$ and the resulting theory is known as the large-energy-effective-theory (LEET) \cite{Charles:1998dr}. Using the HQS for the initial state $B$-meson, and LEET for the final state light-meson, one can factorize the form factors in the soft and hard parts. The soft part of the form factors accounts for the soft gluon interaction with the spectator while the hard-spectator interactions are carried by gluons having a momentum of the order of $m_{B} \Lambda_{QCD}$, where $m_{B}$ is the mass of initial state $B$-meson. These contributions are not independent of each other and in case of $B\to \rho (K^{*}) \ell^{+}\ell^{-}$ decays, it is shown that in the LEET the seven form factors reduced to two in the large recoil limit \cite{Beneke:2000wa}. However, these symmetry relations are not exact and are broken by the radiative corrections. These symmetry-breaking corrections at first order in the strong coupling constant $\alpha_{s}$ are computed by Beneke and Feldmann \cite{Beneke:2000wa} along with their implications to the forward-backward asymmetry in $B\to \rho \ell^{+}\ell^{-}$ decays. Later, these radiative corrections for $B\rightarrow K^* l^+l^-$ are calculated in ref. \cite{Beneke:2001at}.

A close akin of the FCNC mediated $B\to K^{*} \ell^{+}\ell^{-}$ decay is $B\to K_1(1270,1400) \ell^{+}\ell^{-}$, where $K_1(1270,1400)$ are the axial-vector mesons. These axial-vector states are the mixture of the members of two axial-vector $SU(3)$ octet $^3 P_1$ and $^1 P_1$ states, $K_{1A}$ and $K_{1B}$, respectively. 
The physical states $K_1 (1270)$ and $K_1 (1400)$ are related to flavor states $K_{1A}$ and $K_{1B}$ as 
\begin{equation}
  \left( {\begin{array}{cc}
   |K_1 (1270)\rangle  \\
   |K_1 (1400)\rangle  \\
  \end{array} } \right)= 
  \left( {\begin{array}{cc}
   \sin\theta_{K_1} & \cos\theta_{K_1} \\
   \cos\theta_{K_1} & -\sin\theta_{K_1}\\
  \end{array} } \right)
\left( {\begin{array}{cccc}
   |K_{1,A}\rangle  \\
   |K_{1,B}\rangle \\
  \end{array} } \right),\label{mix-states}
\end{equation} 
where $\theta_{K_1}$ is the mixing angle and its value estimated from the radiative $B \to K_1(1270)\gamma$ and $\tau \to K_1(1270)\nu_\tau$ decays is $-(34\pm 13)^{o}$ \cite{Hatanaka:2008}. It is worth emphasizing at this point that the above unitary matrix is also used to relate all the parameters of the $K_{1A,B}$ and physical $K_1(1270,1400)$ states. 

Now, being mediated through the quark level transition, $b\to s \ell^{+}\ell^{-}$, the effective Hamiltonian remains the same in $B\to K^{*} \ell^{+}\ell^{-}$ and $B\to K_1(1270,1400) \ell^{+}\ell^{-}$ decays. Hence, the constraints on the Wilson coefficients and other parameters obtained by analyzing different new physics (NP) scenarios in $B\to K^{*} \ell^{+}\ell^{-}$ can also be used to find the complimentary information from $B\to K_1(1270,1400) \ell^{+}\ell^{-}$ decays. Due to this fact, a thorough analysis of this decay has been done in different NP scenarios (see e.g.,  \cite{Paracha:2007yx, Li:2011nf, Ju:2014oha, Falahati:2014yba, Momeni:2018udf, Momeni:2018tjf, Huang:2018} and references therein). Despite, rigorous NP studies in $B\to K_1(1270,1400) \ell^{+}\ell^{-}$ decays, the contributions arising from the symmetry breaking corrections to the form factors are still missing in the literature and the main motivation of the present study is to fill this gap. In order to achieve this goal, we follow a factorization scheme developed in \cite{Beneke:2000wa} that factorize the soft and hard contributions of the form factors in the framework of the LEET. The corrections to the soft part are manifested in the Wilson coefficients at an order $\alpha_s$ by matching the LEET calculation with the full one-loop QCD calculation.  While for the hard-spectator part, non-perturbative method is required for which we use light cone distribution amplitudes (LCDA). These hard-spectator corrections actually break the symmetry relations. At large recoil, a significant energy is taken by the final state leading light quark for which an expansion over energy is a viable approach. For a more probable final meson state, in which both the leading light and spectator quarks share similar momenta; hard gluon interactions becomes more and more important. The calculation of hard-spectator corrections along with the vertex renormalization in $B\rightarrow K_1(1270)$ is the main objective of this study.  After quantifying these corrections, their impact on the physical observables that are known to have less dependence on the form factors, namely the zero-position of the forward-backward asymmetry and the longitudinal lepton polarization asymmetry, are studied for $B\rightarrow K_1(1270)\mu^+\mu^-$. The case when we have the final state meson to be $K_1(1400)$ is rather trivial from it.

The paper is organized as follows: In section \ref{formfacts}, after a brief introduction of the LEET, its Lagrangian will be given by keeping the final meson mass terms which respect HQS. The seven form factors for $B\rightarrow K_1$, where from here onwards $K_1$ refer to $K_1(1270)$, transition are shown to be written in terms of the two soft form factors $\xi^{\perp,\parallel}_{K_1}(E_F)$ using LEET symmetries. In section \ref{SBCs}, we describe the factorization scheme used to calculate the symmetry breaking corrections to the form factors. The vertex renormalization is carried out along with the hard-spectator interactions to calculate the symmetry breaking  corrections at an order $\alpha_s$. The major uncertainties in the calculation of the form factors lie in hard-spectator corrections especially in the $B$-meson distribution amplitudes. This is discussed in Sec. \ref{forms}. Using light-cone sum rules (LCSR) form factors \cite{Yang:2008zt, Hatanaka:2008ha}, both without corrections and by incorporating the radiative corrections are discussed in the same section. Later their impact on the forward-backward and the longitudinal lepton polarization asymmetries is studied in Sec. \ref{applications}. We conclude in Sec. \ref{conclusion}. The study presented here is supplemented with three appendices: Appendix A summarize the $B$- and $K_1$-mesons distribution amplitudes and the appendix B presents the expressions of different helicity amplitudes for the decay under consideration. Finally, the appendix C gives the detailed calculation of the hard-spectator correction to one of the the form factors $V_2(q^2)$.

\section{Form Factors at Large Recoil}\label{formfacts}
The matrix elements for the decay of $B$-meson to an axial-vector meson $(K_1)$, can be written as;
\begin{eqnarray}
 \left\langle K_{1}(p',\varepsilon^{*})|\bar{q}\gamma^{\mu}b|\bar{B}(p)\right\rangle &=& 2m_{K_1} V_0(q^2)\frac{\varepsilon^*\cdot q}{q^2}q^{\mu}
+(m_B+ m_{K_1})V_1(q^2) 
\left\lbrace\varepsilon^{*\mu} -\frac{\varepsilon^{*}\cdot q}{q^2}q^{\mu}\right\rbrace\notag\\
&&-V_2(q^2)\frac{\varepsilon^{*}\cdot q}{m_B+m_{K_1}}\left\lbrace (p+p')^{\mu}-\frac{m_B ^2-m_{K_1} ^2}{q^2}q^{\mu}\right\rbrace\notag\\
\left\langle K_{1}(p',\varepsilon^{*})|\bar{q}\gamma^{\mu}\gamma_{5}b|\bar{B}(p)\right\rangle &=&\frac{2i A(q^2)}{m_B+m_{K_1}}\epsilon^{\mu\nu\rho\sigma}\varepsilon^{*} _{\nu} p'_{\rho}p_{\sigma}\notag\\
\left\langle K_{1}(p',\varepsilon^{*})|\bar{q}\sigma^{\mu\nu}q_{\nu}\gamma_{5}b|\bar{B}(p)\right\rangle &=&-2 T_1(q^2)\epsilon^{\mu\nu\rho\sigma}\varepsilon^{*} _{\nu} p'_{\rho}p_{\sigma}\notag\\
\left\langle K_{1}(p',\varepsilon^{*})|\bar{q}\sigma^{\mu\nu}q_{\nu}b|\bar{B}(p)\right\rangle &=&-i T_2(q^2)\left\lbrace(m_B^2-m_{K_1} ^2)\varepsilon^{*\mu} -(\varepsilon^*\cdot q)(p +p')^{\mu}\right\rbrace\notag\\
&&-i T_3(q^2)(\varepsilon^{*} \cdot q)\left\lbrace q^\mu-\frac{q^2}{m_B^2-m_{K_1} ^2}(p +p')^{\mu}\right\rbrace\label{mat-Ele1}
\end{eqnarray} 
where $p^{\mu}(p'^{\mu})$, $m_B(m_{K_1})$ are the momentum, mass of $B(K_1)$-meson, respectively, and $\varepsilon^{*\mu}$ is the polarization vector of the $K_1$-meson. 
The interaction of heavy quarks with soft gluons render the heavy quark effective theory (HQET) \cite{Neubert:1993mb}. Using $q^2=(p-p')^2$, the energy of the final state $K_1$- meson is $E_F = (m_B^2+m_{K_1} ^2 -q^2)/2m_B$. In the large recoil region, the momentum transfer squared, i.e., $q^2$ is small and because of the fact that $m_{K_1} ^2<<m_B^2$ and $q^2<<m_B^2$, the energy of $K_1$-meson is of order $E_F\sim m_B/2$. Therefore, the $E_F$ is a good expansion parameter and for the $K_1$-meson whose mass is larger than $1$ GeV; retaining terms of the order of $m_{K_1} ^2/m_B^2$ would be interesting. However, it is safe to neglect the terms of the order of  $\Lambda_{QCD}/E_F$ .

The four-momentum of $B-$meson in terms of the velocity $v$ of the heavy quark can be expressed as $p^{\mu} =m_B v^{\mu}$.
Being a bound state of a heavy and a light quark, the heavy quark has momentum
\begin{equation}
p_Q ^\mu = m_Q v^\mu + k^\mu ,\label{PQ}
\end{equation}
where $k$ is the residual momentum and $|k|\sim \Lambda_{QCD}<< m_b$. In $B$-meson rest frame, the components of the four velocity are $v=(1,0,0,0)$. In order to work with light-cone variables, let us introduce two light like vectors $n_\mu$ and $\eta_\mu$ satisfying $n^2=\eta^2=0$. Choosing $n=(1,0,0,1)$ and $\eta=(1,0,0,-1)$ it can be verified that $\eta_\mu = 2v_\mu -n_\mu$ while $n\cdot v=1$ and $n\cdot\eta =2$.

The momentum of $K_1$-meson in terms of $n_\mu$ and $\eta_\mu$ can be expressed as \cite{Ebert:2001pc}
\begin{equation}
p_F = En+\frac{m_{K_1} ^2}{4E}\eta, \label{PF}
\end{equation}
with $E$ being the off-shell energy. The on-shell energy and $3-$momentum of the final state $K_1$-meson are
\begin{equation}
E_F =E\left(1+\frac{m_{K_1} ^2}{4E^2}\right),\quad\quad \Delta\equiv\sqrt{E_F ^2 -m_{K_1} ^2}=E\left(1-\frac{m_{K_1} ^2}{4E^2}\right) \label{EF}
\end{equation}
and the momentum of the leading light quark in final state is given as
\begin{equation}
p_q =E n+\frac{m_{K_1} ^2}{4E}\eta +k'=\Delta n+\frac{m_{K_1} ^2}{4E}v+k',\label{pqq}
\end{equation}
where again the residual momentum is of the order of $|k'|\sim\Lambda_{QCD}<<E$. The effective Lagrangian can be derived from the $4-$component QCD quark fields $q(x)$. The two-component light-quark fields are given via projection operators $(P_{\pm})$ as
\begin{equation}
q_{\pm}(x)=\exp\left(i \Delta\cdot x+\frac{m_q^2}{2E}v\cdot x\right)P_{\pm}q(x), \label{qpmx}
\end{equation}
with 
\begin{eqnarray}
P_{+} &=& \frac{\slashed{n}\slashed{v}}{2}, \qquad P_+ ^2=P_+ ,\notag\\
P_{-} &=&\frac{\slashed{v}\slashed{n}}{2},\qquad P_- ^2=P_-. \label{projection}
\end{eqnarray} 
The effective Lagrangian up-to order $\Lambda_{QCD}/E$ while retaining terms of order $m_{K_1} ^2/E$ can be expressed by
\begin{equation}
\mathcal{L} = \bar{q}(x)\slashed{v}\left(i n\cdot D+\frac{m_{K_1} ^2}{2E}\right) q(x), \label{lq-Lag}
\end{equation}
where $D^\mu =\partial^\mu -ig_s A^\mu$ is the covariant derivative. The second term in the Lagrangian gives the contribution due to final state meson mass and it does not break the symmetry of the leading order Lagrangian due to similar Dirac structure. In case of the vector-meson $(\rho, K^*)$, whose mass is less then 1 GeV, the terms of the order $\frac{m^2_{K^*}}{2E}$ are ignored in $B\to (\rho,\; K^*)\ell^{+}\ell^{-}$ calculations \cite{Beneke:2000wa, Beneke:2001at}.

In our calculation of form factors of $B-$meson decaying to $K_{1}$-meson, it is instructive to define the sum and difference of momenta
\begin{eqnarray}
(p+p')^\mu &=& m_B\left(1+\frac{m_{K_1} ^2}{2m_BE}\right)v^\mu+\Delta n^\mu,\notag \\
q\equiv (p-p')^\mu&=& m_B\left(1-\frac{m_{K_1} ^2}{2m_BE}\right)v^\mu -\Delta n^\mu \label{mom-trans}.
\end{eqnarray}
The $K_1$-meson with polarization vector $\varepsilon^{*\mu}$ satisfy the transverse relation, i.e., $\varepsilon^*\cdot p'=0$. From Eq. (\ref{mom-trans}), contracting with $\varepsilon^{*\mu}$ and making use of transverse relation, we have a useful identity
\begin{equation}
\varepsilon^*\cdot n= -\frac{m_{K_1} ^2}{2E\Delta}(\varepsilon^*\cdot v) \label{identity}.
\end{equation} 
Using the technique familiar from the HQET, the soft form factors relation can be found as \cite{Neubert:1993mb}
\begin{equation}
\langle K_{1}(p',\varepsilon^{*})|\bar{q}\Gamma b|B(p)\rangle =\text{Tr}[A(E_F)\bar{\mathit{M}}_{K_1}\Gamma\mathit{M}_B], \label{BtoA-1}
\end{equation}
where the projector for $K_1$-meson is defined as $\bar{\mathit{M}}_{K_1}=-\varepsilon^* \gamma_5 \frac{\slashed{v}\slashed{n}}{2}$ and for $B$-meson it is $\mathit{M_B}=-\frac{1+\slashed{v}}{2}\gamma_5$.  The function $A(E_F)$ contains the long distance dynamics that is independent of any Dirac structure $\Gamma$ and it can be written as
\begin{equation}
A(E_F)= E_F \slashed{n}\left\lbrace \xi^\perp _{K_1}(E_F)-\frac{\slashed{v}}{2}\xi^\parallel _{K_1}(E_F)\right\rbrace ,
\end{equation}
with $\xi^\perp_{K_1} (E_F)$ and $\xi^{\parallel}_{K_1}(E_F)$ denote the contribution to form factors of transversely and longitudinally polarized $K_1$-meson, respectively. Substituting $\mathit{M}_{K_1},\mathit{M}_B$ along with the function $A(E_F)$ in Eq. (\ref{BtoA-1}) and by considering the possible Dirac structures $\Gamma = \lbrace\gamma^\mu,\gamma^\mu\gamma_5,\sigma^{\mu\nu} q_\nu ,\sigma^{\mu\nu}q_\nu \gamma_5\rbrace$, the trace calculation gives
\begin{eqnarray}
\langle K_{1}(p',\varepsilon^{*})|\bar{q}\gamma^{\mu}b|B(p)\rangle &=&2E_F \xi^{\perp} _{K_1}(E_F)\left[\varepsilon^{*\mu}-(\varepsilon^{*}\cdot v)\left(\frac{E_F}{\Delta}n^{\mu}-\frac{m_{K_1} ^2}{2E\Delta}v^\mu\right)\right] \notag\\
&&+2E_F\left(1+\frac{m^2 _{K_1}}{4E\Delta}\right)\xi^{\parallel} _{K_1}(E_F)(\varepsilon^{*}\cdot v) n^\mu , \notag\\
\langle K_{1}(p',\varepsilon^{*})|\bar{q}\gamma^{\mu}\gamma_5 b|B(p)\rangle &=&2iE_F \xi^{\perp} _{K_1}(E_F) \varepsilon^{\mu\nu\rho\sigma}\varepsilon^{*} _{\nu}n_{\rho}v_{\sigma} ,\notag\\
\langle K_{1}(p',\varepsilon^{*})|\bar{q}\sigma^{\mu\nu}\gamma_5 q_\nu b|B(p)\rangle &=&2E_F m_B\xi_{K_1} ^{\perp}\left(1-\frac{m_{K_1} ^2}{2Em_B}\right)\varepsilon^{\mu\nu\rho\sigma}\varepsilon^{*} _{\nu}n_{\rho}v_{\sigma} ,\notag\\
\langle K_{1}(p',\varepsilon^{*})|\bar{q}\sigma^{\mu\nu}q_\nu b|B(p)\rangle &=&2iE_F\left[ m_B\xi^{\perp} _{K_1}(E_F)\left(1-\frac{m_{K_1} ^2}{2E m_B}\right) \left\lbrace\varepsilon^{* \mu}-\varepsilon^{*}\cdot v\left(\frac{E_F}{\Delta}n^{\mu}-\frac{m_{K_1} ^2}{2E\Delta}v^\mu\right)\right\rbrace\right. \notag\\
&&\left.+\frac{E}{\Delta}\xi^{\parallel} _{K_1}(E_F)(\varepsilon^{*}\cdot v) \left\lbrace(m_B-E_F)n^{\mu}-m_B\left(1-\frac{m_{K_1} ^2}{2Em_B}\right)v^{\mu} \right\rbrace\right].\label{BtoA-2}
\end{eqnarray}
 
It is important to emphasis that despite the similarity that both $B \to K^{*}\ell^{+}\ell^{-}$ and $B \to K_{1}\ell^{+}\ell^{-}$ decays are mediated by the quark level transition $b \to s\ell^{+}\ell^{-}$, there are some differences. The first and the obvious difference is that $K_1$ is an axial-vector meson and due to this fact, the matrix elements corresponding to vector and axial-vector currents in $K^{*}$ case are interchanged here and this can be seen in Eq. (\ref{mat-Ele1}). The second common difference between $K^{*}$  and $K_1(1270,1400)$ is that the later states are a mixture of flavor eigenstates $K_{1A,1B}$ and hence the corresponding form factors and other quantities will also mix which is not the case for $K^{*}$ meson. The last and the most particular one is that contrary to the $K^{*}$ meson, the mass of $K_1-$meson is above 1 GeV and hence without ignoring its mass such symmetry relations for form factors were earlier calculated in \cite{Ebert:2001pc}. Therefore, we have also kept the $K_1$ meson mass terms in calculating vertex and hard-spectator corrections. By ignoring the final state meson mass in these correction terms and also interchanging the role of vector and axial-vector currents, one can see that the corresponding relations for the $B \to K^{*}\ell^{+}\ell^{-}$ can be reproduced. 

Now comparing Eq. (\ref{mat-Ele1}) with Eq. (\ref{BtoA-2}) to represent all seven form factors in terms of the two soft form factors $\xi^{\perp,\parallel}_{K_1}(E_F)$, one gets
\begin{eqnarray}
V_0(q^2)&=&\frac{E_F}{m_{K_1}}\xi^{\parallel}_{K_1}(E_F),\label{V0}\\
V_1(q^2)&=&\frac{2E_F}{m_B+m_{K_1}}\xi^\perp _{K_1}(E_F),\label{V1} \\
V_2(q^2)&=&\left(1+\frac{m_{K_1}}{{m_B}}\right)\left(1+\frac{2m_{K_1}^2}{m_B^2}\right)\left[\xi^\perp _{K_1}(E_F) - \xi^{\parallel}_{K_1}(E_F)\right],\label{V2}\\
A(q^2) &=&\left(1+\frac{m_{K_1}}{{m_B}}\right)\frac{E_F}{\Delta}\xi^{\perp}_{K_1}(E_F),\label{Aq}\\
T_1(q^2)&=&\left(1-\frac{m^2 _{K_1}}{m_B^2}\right)\frac{E_F}{\Delta}\xi^{\perp}_{K_1}(E_F),\label{T1}\\
T_2 (q^2)&=&\frac{2E_F}{m_B}\xi^{\perp}_{K_1}(E_F),\label{T2}\\
T_3(q^2)&=&\left(1+\frac{5m_{K_1} ^2}{m_B^2}\right)\xi^{\perp}_{K_1}(E_F)-\left(1+\frac{2m_{K_1} ^2}{m_B^2}\right) \xi^{\parallel}_{K_1}(E_F).\label{T3}
\end{eqnarray}
Recall  that the energy $E_F$ is a function of $q^2$ and for the radiative decays  $q^2=0$, therefore, we get the trivial expressions for the form factors. The form factor $V_0 (q^2)$ only depends upon $\xi^{\parallel}_{K_1}(E_F)$ and it will be later used as a renormalization convention along with $A(q^2)$ for perpendicular-polarization form factor $\xi^{\perp}_{K_1}(E_F)$.
 
\section{Symmetry breaking corrections}\label{SBCs}
\begin{figure}
  \includegraphics[width=75mm]{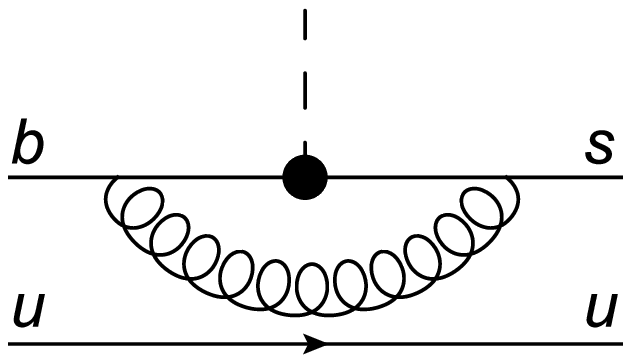}
  \caption{Vertex correction in $B\to K_1$ decays.}
  \label{fig:FIGURE1}
\end{figure}

\begin{figure}
  \includegraphics[width=\linewidth]{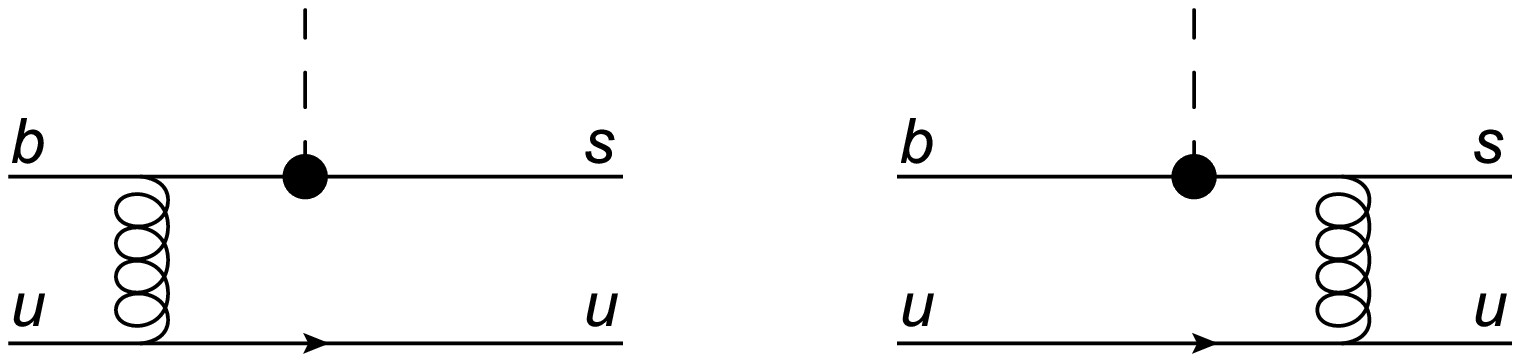}
  \caption{Hard spectator corrections to the form factors in $B\to K_1$ decays.}
  \label{fig:FIGURE2}
\end{figure}

The symmetries arise in the HQET help us to relate the form factors and hence reduce the number of independent form factors e.g., from seven to two in the decay under consideration. However, the heavy quark/large recoil symmetries are broken by the radiative corrections. These corrections  arise from vertex diagram as shown in Fig.\ref{fig:FIGURE1} or from the hard-spectator scattering diagrams shown in Fig. \ref{fig:FIGURE2}. For HQS at large recoil, we worked out the relations of the soft form factors in section \ref{formfacts}. The vertex contribution arise at the order of $1/m_B$ and $\alpha_s$. The contributions arising from hard-spectator interactions are suppressed by an order of $\alpha_s$. However, in case of heavy-to-light transitions, these corrections are still important. This is due to the fact that we desire a probable configuration in which the momentum of spectator and leading quark scale in a similar fashion. This requires a well thought factorization scheme and to serve this purpose, there had been few factorization schemes developed in the last ten-to-fifteen years for these heavy-to-light transitions (e.g.  \cite{Bauer:2002aj, Beneke:2000wa}). Beneke and Feldmann \cite{Beneke:2000wa} have developed a factorization scheme in terms of soft- and hard-contributions to the form factors in the framework of LEET and we adopt it in our present work. We will see that the vertex corrections do not respect the symmetry relations and to take them into account, these are calculated at an extra order of  $\alpha_s$ in an effective theory and then matched with full theory. The difference in matching the two will give us the required contributions. There is one subtlety, LEET is not infrared safe, because it does not take care of the  collinear gluons and hence can not correctly reproduce infrared divergences. However, this will be taken care of in the future work \cite{AJSI}. But in context of this work, such collinear gluons do not break symmetry relations among the soft form factors \cite{Bauer:2000yr}. Similarly, for the hard-spectator scattering, it is seen that end-point divergences respect the HQS and are also accounted for in soft-form factors. We would like to mention here that the hard gluon has virtuality of $m_B\Lambda_{QCD}$ while the energy of the leading light quark scales as $m_B/2$. To the form factors defined in Eqs. (\ref{mat-Ele1}) the factorization formula for heavy to light case at leading order in $1/m_B$ reads as  \cite{Beneke:2000wa}
\begin{equation}
f_i(q^2)=C_i \xi^{a}_{K_1}(E_F)+\Phi_B \otimes\mathcal{T}^\Gamma\otimes\Phi_{K_1},\label{factorization-sch}
\end{equation}
where $\Phi_B$ and $\Phi_{K_1}$ are the light-cone distribution amplitudes for the $B$- and $K_{1}$- mesons, respectively. The first term accounts for the soft contributions with $\xi^{a}_{K_1}(E_F)$ representing the soft form factors and the label $a=\perp ,\parallel$ corresponds to the perpendicular, parallel polarizations of the $K_1$-meson. In sec. \ref{formfacts}, we have already expressed the seven form factors in terms of $\xi^{\perp , \parallel}_{K_1}(E_F)$ (c.f.,  Eq. (\ref{BtoA-2})). The second term in Eq. (\ref{factorization-sch}) depicts the hard-spectator scattering contributions. We will see later, while calculating these corrections; that the corresponding amplitude $\mathcal{T}^{\Gamma}$ has the end-point divergences. These end-point divergences are then absorbed in the soft form factors.
\subsection{Vertex Renormalization}
Making use of the Passarino-Veltman reduction and keeping the mass term for the $K_1$-meson, the vertex diagram shown in Fig. \ref{fig:FIGURE1} can be evaluated in a systematic way. Let's write $\bar{u}(p')$ and $u(p)$ as Dirac spinors for light (assumed to be massless) and heavy quarks and introduce a small mass $\lambda$ for gluon to regulate IR-divergences. The UV-divergences are dealt in a naive dimensional regularization (NDR) scheme ($d=4-2\epsilon$) and utilizing  $\left(\bar{u}(p')\slashed{p'}=0, \slashed{p}u(p)=m_b u(p)\right)$  to get the relation for an arbitrary current $\Gamma$;
\begin{eqnarray}
\bar{u}(p')\Gamma(p',p)u(p)&=&\frac{\alpha_s C_F}{4\pi}\bar{u}(p')\left[\left\lbrace -\frac{1}{2}\ln\left(\frac{\lambda^2 m_b^2}{m_b ^2 -q^2}\right)-2\ln\left(\frac{\lambda^2 m_b ^2}{(m_b ^2 -q^2)^2}\right)-2Li_2 \left(\frac{q^2}{m_b ^2}\right)+\frac{m_B}{E_F}L-3-\frac{\pi^2}{2}\right\rbrace\right.\Gamma \notag\\
&&+\frac{1}{4}\left\lbrace \frac{1}{\hat{\epsilon}}+3-\ln\left(\frac{m_b ^2}{\mu^2}\right)-L'\right\rbrace \gamma^\alpha\gamma^\beta\Gamma\gamma_{\beta}\gamma_{\alpha}
+\frac{1}{2q^2}\left\lbrace 1-\frac{ m_B}{2 E_F}L\right\rbrace \gamma^{\alpha}\slashed{p}\Gamma\slashed{p^{'}}\gamma_{\alpha}\notag\\
&&+\frac{1}{2q^2}\left\lbrace 1-L^{'}\right\rbrace m_b \gamma^{\alpha}\slashed{p}\Gamma \gamma_{\alpha}-\left.\frac{1}{2q^2}\left\lbrace 2-L^{''}\right\rbrace m_b \Gamma \slashed{p'}\right] u(p), \label{current}
\end{eqnarray}
where $q^2=m_B^2+m_{K_1} ^2-2m_B E_F$. We defined the pole $1/\hat{\epsilon}=1/\epsilon -\gamma_E +$ln$4\pi$ which in the $\overline{MS}$ scheme will be subtracted out. The currents are defined in the NDR with an anti-commuting $\gamma_5$.  The remaining quantities are
\begin{eqnarray*}
L&=&-\frac{2E_F}{m_B-2E_F +\frac{m_{K_1} ^2}{m_B}}\ln\left(\frac{2E_F}{ m_B}-\frac{m_{K_1} ^2}{ m_B^2}\right),\nonumber\\
L^{'}&=&L\left(1-\frac{m_{K_1} ^2}{2E_F m_B}\right),\nonumber\\
L^{''}&=&L\left(4-\frac{m_B  }{E_F}-\frac{2m_{K_1} ^2}{E_F m_B}\right).
\end{eqnarray*}
The coefficient $C_i$ at 1-loop order are calculated by finding the difference between full theory and the LEET vertex calculation. It can be seen that all infrared divergent terms in (\ref{current}) have same structure as $\Gamma$ so they can be absorbed in the redefinition of the soft form factors $\xi^{\perp,\parallel}_{K_1}$.  Introducing a renormalization convention for an axial-vector meson form factors that holds exactly to all orders in perturbation theory, we have
\begin{eqnarray}
A(q^2)=\left(1+\frac{m_{K_1}}{{m_B}}\right)\frac{E_F}{\Delta}\xi^\perp_{K_1}(E_F);\qquad\qquad V_0(q^2)=\frac{E_F}{m_{K_1}}\xi^\parallel_{K_1}(E_F).\label{newAq}
\end{eqnarray}
For a given current $\Gamma$ in Eq. (\ref{current}), one can find the $\mathcal{O}(\alpha_s)$ corrections by substituting in  Eq. (\ref{BtoA-1}). Making use of renormalization convention in Eq. (\ref{newAq}) and comparing it with the form factors defined in Eq. (\ref{mat-Ele1}) gives us
\begin{eqnarray}
A(q^2)&=&\left(1+\frac{m_{K_1}}{{m_B}}\right)\frac{E_F}{\Delta}\xi^\perp _{K_1}(E_F),\notag \\
V_0(q^2)&=&\frac{E_F}{m_{K_1}}\xi^\parallel_{K_1}(E_F)\notag\\
V_1(q^2)&=&\frac{2E_F}{m_B+m_{K_1}}\xi^\perp _{K_1}(E_F),\notag\\
V_2(q^2)&=&\left(1+\frac{m_{K_1}}{{m_B}}\right)\frac{E_F}{\Delta}\left[\xi^\perp _{K_1}(E_F)-\left(1+\frac{\alpha_s C_F}{4\pi}(-2+2L')\right)\xi^\parallel_{K_1}(E_F)\right],\notag\\
T_1(q^2)&=&\frac{E_F}{\Delta}\left[\left(1-\frac{m_{K_1} ^2}{m_B^2}\right)+\frac{\alpha_s C_F}{4\pi}\left(-\frac{m_B}{E_F}L+\ln\left(\frac{m_b^2}{\mu^2}\right)+L'\right)\right]\xi^\perp _{K_1}(E_F),\notag
\end{eqnarray}
\begin{eqnarray}
T_2(q^2)&=&\frac{2E_F}{m_B}\left[1+\frac{\alpha_s C_F}{4\pi}\left(\frac{m_B}{E_F}L-\ln\left(\frac{m_b^2}{\mu^2}\right)-L'\right)\right]\xi^\perp _{K_1}(E_F),\label{newff}\\
T_3(q^2)&=&\left(1+\frac{m^{2}_{K_1}}{{m^{2}_B}}\right)\left[\left\lbrace 1+\frac{\alpha_s C_F}{4\pi}\left(\frac{m_B}{E_F}L-\ln\left(\frac{m_b^2}{\mu^2}\right)-L'\right)\xi^\perp _{K_1}(E_F)\right\rbrace\right.\notag\\
&&\left.-\left\lbrace1+\frac{\alpha_s C_F}{4\pi}\left(\frac{3m_B}{2E_F}L -\ln\left(\frac{m_b^2}{\mu^2}\right)+L'-2\right)\xi^\parallel_{K_1}(E_F)\right\rbrace\right].\notag
\end{eqnarray} 
Here, one can notice that $V_1(q^2)$ does not receive any contribution from the vertex corrections. 

\subsection{Hard-Spectator Corrections}

The hard-spectator corrections arise at an order $\alpha_s$ for which the two-particle light-cone distribution amplitudes of the $B$- and the light $K_1$-mesons are given in Appendix \ref{AppendixA}. The momenta of $b$-quark and spectator quark before the decay are
\begin{equation}
 p ^\mu =m_b v^\mu , \quad\quad\quad l^\mu = \frac{l_+}{2}n_+ ^\mu +l_\perp ^\mu+\frac{l_-}{2}n_- ^\mu. \label{B-momenta}
\end{equation}
After the quark level transition $b\rightarrow s$ which governs the $B\to K_1$ decay, the momenta of the leading $s$-quark and the corresponding spectator quark in the $K_1$-meson are
\begin{eqnarray}
 k_1 ^\mu &=& uE_F n_- ^\mu +k_\perp ^\mu +\left(\frac{\vec{k}^2 _\perp}{4uE_F} +\frac{m_{K_1} ^2}{4u E_F}\right)n_+ ^\mu,\label{k1momenta}\\
 k_2 ^\mu &=&\bar{u}E_F n_- ^\mu -k_\perp ^\mu +\left(\frac{\vec{k}^2 _\perp}{4\bar{u}E_F}+\frac{m_{K_1} ^2}{4\bar{u} E_F}\right)n_+ ^\mu, \label{gluon-momenta}
\end{eqnarray}
where $\bar{u}=1-u$. All components of $l$ as well as $k_\perp$ in $k_{1,2}$ are of the order of $\Lambda_{QCD}$ . It can be seen that $(k_1 + k_2)^2\sim$ $m_{K_1} ^2$ which otherwise was scaling like $\Lambda_{QCD} ^2$. We are interested in hard-exchanges where gluon momenta scale as $m_B\Lambda_{QCD}$, therefore terms proportional to $\slashed{n}_-$ will matter.

The contributions to heavy-to-light matrix elements for a given current is given by the convolution formula
\begin{equation}
\langle K_1|\bar{q}\Gamma b|B\rangle =\frac{4\pi\alpha_s C_F}{N_C}\int_0 ^1 du\int_0 ^\infty dl_+ \mathcal{M}_{jk} ^B \mathcal{M}_{li} ^{K_1} \mathcal{T}_{ijkl} ^\Gamma \label{matrix-elems},
\end{equation}
where $\mathcal{M}^B, \mathcal{M}^{K_1}$ are two-particle light-cone projectors which contain the non-perturbative bound state dynamics. $\mathcal{T}^\Gamma$ is the hard scattering amplitude which is calculated from the Feynman diagrams in Fig. \ref{fig:FIGURE2}. The $K_1$-meson projector is given as
\begin{equation}
\mathcal{M}^{K_1} = \left[-\frac{i}{4}\left\lbrace f_{K_1} ^{\perp}\slashed{\varepsilon}^*\slashed{p'}\phi^{K_1}_{\perp}(u)+f_{K_1} ^{\parallel} \frac{m_{K_1}}{E}(v\cdot\varepsilon^*)\phi^{K_1}_{\parallel}(u)\right\rbrace\gamma_5\right]_{li}\label{K1-projectors}
\end{equation}
with $f_{K_1} ^{\perp}$ and $f_{K_1} ^{\parallel}$ denoting the transverse and longitudinal vector meson decay constants and $\phi^{K_1}_{\perp, \parallel}$ denote the twist-3 two-particle distribution amplitudes (c.f. Appendix  \ref{AppendixA})

 Similarly the projector for $B$-meson is
\begin{equation}
\qquad \mathcal{M}^B =-\left.\frac{if_B m_B}{4}\left[\frac{1+\slashed{v}}{2}\left\lbrace\phi^B _+ (l_+)\slashed{n}_+ +\phi_- ^B (l_+) \left(\slashed{n}_- -l_+ \gamma_{\perp}^v\frac{\partial}{\partial l_\perp ^v}\right)\right\rbrace\gamma_5\right]_{jk}\right|_{l=(l_+ /2)n_+}\label{B-projectors}
\end{equation}
where $f_B$ is the $B$-meson decay constant. Last but not least the hard-scattering amplitude from the Fig. \ref{fig:FIGURE2} is
\begin{equation}
\mathcal{T}^\Gamma _{ijkl}=\left[\Gamma \frac{m_b(1+\slashed{v})+\slashed{l}-\slashed{k_2}}{(m_b v +l -k_2)^2-m_b^2}\gamma_\mu +\gamma_\mu \frac{\slashed{k_1}+\slashed{k_2} -\slashed{l}}{(k_1 +k_2 -l)^2}\Gamma\right]_{ij}\frac{1}{(l-k_2)^2}[\gamma^\mu]_{kl}.\label{Tmatrix}
\end{equation}
Working in the context of the LEET, we are interested only in the hard gluon exchange. The gluon propagator with momenta defined in Eqs. (\ref{B-momenta}, \ref{k1momenta}, \ref{gluon-momenta}) when expanded gives $(l-k_2)^2\sim-2l_+ \bar{u}E_F$ which scales as $m_B\Lambda_{QCD}$. The first term in Eq. (\ref{Tmatrix}) has numerator  $\slashed{l}-\slashed{k_2}\sim -\bar{u}E_F\slashed{n}_-$ that is the surviving term at the scale $m_B\Lambda_{QCD}$. When all the dust settle down, the contributions of hard exchanges after neglecting terms of order $\Lambda_{QCD}/m_B$ are
\begin{equation}
\mathcal{T}^\Gamma _{ijkl}\simeq\left[\Gamma \frac{m_b(1+\slashed{v})-\bar{u}E_F \slashed{n}_-}{4\bar{u}^2 l_+ m_b E_F ^2}\gamma_\mu +\gamma_\mu \frac{E_F\slashed{n}_- -\slashed{l}}{4\bar{u}l_+ ^2 E_F^2}\Gamma\right]_{ij}[\gamma^\mu]_{kl}. \label{Tmatrix1}
\end{equation}
The term $m_b(1+\slashed{v})$ diverges logarithmically for $\bar{u}\rightarrow 0$ as the functions $\phi(u)$ vanishes only linearly in the leading twist. These so called end-point divergences can be absorbed in soft form factors in our factorization scheme (\ref{factorization-sch}) as they do not break heavy/large recoil symmetry relations. This can be easily verified by looking at the similarity of its current structure with the one defined in Eq. (\ref{BtoA-1}). In this study, we can go beyond the twist-2 and work with the twist-3 distribution amplitudes. Now the twist-3 distributions are suppressed by $1/m_B$ which are compensated by the linear term $m_B/\Lambda_{QCD}$ in $\mathcal{M}^{K_1}$ for the case $\bar{u}\rightarrow 0$. Therefore, such terms do contribute at leading order in soft factors $\xi^{\perp ,\parallel}_{K_1}$. Similar argument can be made for second term in Eq. (\ref{Tmatrix1}) where the twist-3 contributions are factored in soft form factors for the case $\bar{u}\rightarrow 0$ as end-point divergences. In short, all the twist-3 contributions can be seen to preserve heavy quark/large-energy symmetries. There remains just one term, i.e,  $-\bar{u}E_F \slashed{n}_-$ in first numerator of Eq. (\ref{Tmatrix1}) which breaks the symmetry. This term is linear in $\bar{u}$ and after evaluating convloution formula in  Eq. (\ref{matrix-elems}), term proportional to $\phi_+ ^B ({l_+)}$ survive. Now, let us focus on $\phi_- ^B (l_-)$ in $\mathcal{M}^B$ given in Eq. (\ref{B-projectors}).  The $\slashed{n}_-$ from the term $(-\bar{u}E_F\slashed{n}_-)$ when multiplied with $\phi_- ^B(l_-)\left(\slashed{n}_- - l_+ \gamma_\perp ^\nu \frac{\partial}{\partial l_\perp ^\nu}\right)$ gives a term proportional to $l_+ \gamma_\perp$ as $\slashed{n}_- \slashed{n}_- =0$. This remaining term involving $\phi_- ^B(l_-)$ can be found to preserve heavy quark/large recoil symmetry. 

Just to give some details of how the symmetry breaking correction terms appear in the form factors, the detailed calculation of the hard-spectator corrections to the form factor $V_2(q^2)$ is given in Appendix \ref{AppendixC}. Similar procedure is adopted for the form factors $T_{1,2,3}(q^2)$. Now using different possible Dirac gamma structures in Eq. (\ref{matrix-elems}) and comparing it with Eq. (\ref{mat-Ele1}), the complete results of form factors after including vertex and hard-spectator corrections become
\begin{eqnarray}
A(q^2)&=&\left(1+\frac{m_{K_1}}{m_B}\right)\frac{E_F}{\Delta}\xi^{\perp}_{K_1}(q^2),\label{Aqff}\\
V_0(q^2)&=&\frac{E_F}{m_{K_1}}\xi^{\parallel}_{K_1}(q^2),\label{V0ff}\\
V_1(q^2)&=&\frac{2E_F}{m_B+m_{K_1}}\xi^\perp_{K_1}(q^2),\label{V1ff}\\
V_2(q^2)&=&\left(1+\frac{m_{K_1}}{m_B}\right)\frac{E_F}{\Delta}\left[\xi^{\perp}_{K_1}(q^2) -\left\lbrace 1+\frac{\alpha_s C_F}{4\pi}\left(-2+2L'-\frac{q^2}{m_B^2-m_{K_1} ^2}\frac{m_B}{2E_F}\frac{\bold{\Delta} F_{\parallel}}{\xi^{\parallel}(q^2)}\right)\right\rbrace\xi^{\parallel}_{K_1}(q^2)\right],\label{V2ff}\\
T_1(q^2)&=&\frac{E_F}{\Delta}\left[(1-\frac{m_{K_1} ^2}{m_B^2})+\frac{\alpha_s C_F}{4\pi}\left\lbrace -\frac{ m_B}{E_F}L+\ln\left(\frac{m_b ^2}{\mu^2}\right)+L'+\frac{\Delta m_B}{4E_F ^2}\frac{\bold{\Delta} F_{\perp}}{\xi^{\perp}_{K_1}(q^2)}\right\rbrace\right]\xi^{\perp}_{K_1}(q^2),\label{T1ff}\\
T_2(q^2)&=&\frac{2E_F}{m_B}\left[1+\frac{\alpha_s C_F}{4\pi}\left\lbrace\frac{- m_B}{E_F}L+\ln\left(\frac{m_b ^2}{\mu^2}\right)+L'+\frac{m_B}{4E_F}\frac{\bold{\Delta}F_\perp}{\xi^\perp_{K_1}(q^2)}\right\rbrace\right]\xi^{\perp}_{K_1}(q^2),\label{T2ff}\\
T_3(q^2)&=&\left(1-\frac{m_{K_1} ^2}{m_B^2}\right)\left[\left\lbrace 1+\frac{\alpha_s C_F}{4\pi}\left(\frac{m_B}{E_F}L-\ln\left(\frac{m_b^2}{\mu^2}\right)-L'-\frac{m_B^2}{m_B^2-m_{K_1} ^2}\frac{m_B}{4E_F}\frac{\bold{\Delta} F_\perp}{\xi^\perp_{K_1}(q^2)}\right)\right\rbrace\xi^\perp_{K_1}(q^2)\right.\notag\\
&&\left. -\left\lbrace 1+\frac{\alpha_s C_F}{4\pi}\left(\frac{3m_BL}{2E_F} -\ln\left(\frac{m_b^2}{\mu^2}\right)+L'-2\right\rbrace\xi^\parallel_{K_1}(q^2)\right) \right]\label{T3ff}.
\end{eqnarray}
where the quantities, $\bold{\Delta} F_{\perp}$ and $\bold{\Delta} F_{\parallel}$ can be written as
\begin{eqnarray}
\bold{\Delta} F^{\perp}&=&\frac{8\pi^2 f_B f^\perp _{K_1}}{N_C m_B}\langle l_+ ^{-1}\rangle_+ \langle\bar{u}^{-1}\rangle^\perp,\notag\\
\bold{\Delta} F^{\parallel}&=&\frac{8\pi^2 f_B f_{K_1}^\parallel}{N_C m_B}\langle l_+ ^{-1}\rangle_+ \langle\bar{u}^{-1}\rangle^\parallel.
\end{eqnarray}
The leading twist moments given for $K_1$- and $B$- meson are
\begin{equation}
\langle\bar{u}^{-1}\rangle^{\perp,\parallel}=\int du \frac{\phi^{K_1} _{\perp,\parallel}(u)}{\bar{u}}\label{phib},
\end{equation}
and
\begin{equation}
\langle l_+ ^{-1}\rangle_{+}=\int dl_+ \frac{\phi^{B}_+ (l_+)}{l_+}\label{phil}.
\end{equation}
Again, we can see that $V_1(q^2)$ does not receive any symmetry breaking correction term, i.e., neither from vertex and nor from hard-spectator corrections. 
\begin{figure}
\includegraphics{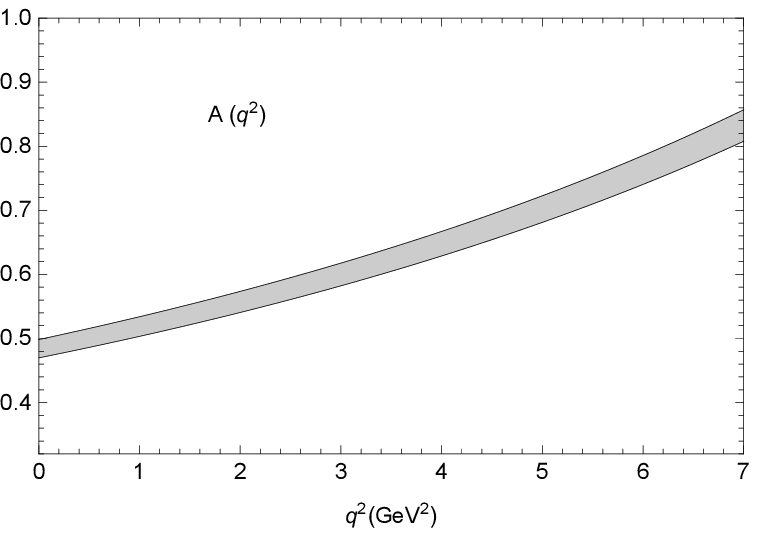}
\includegraphics{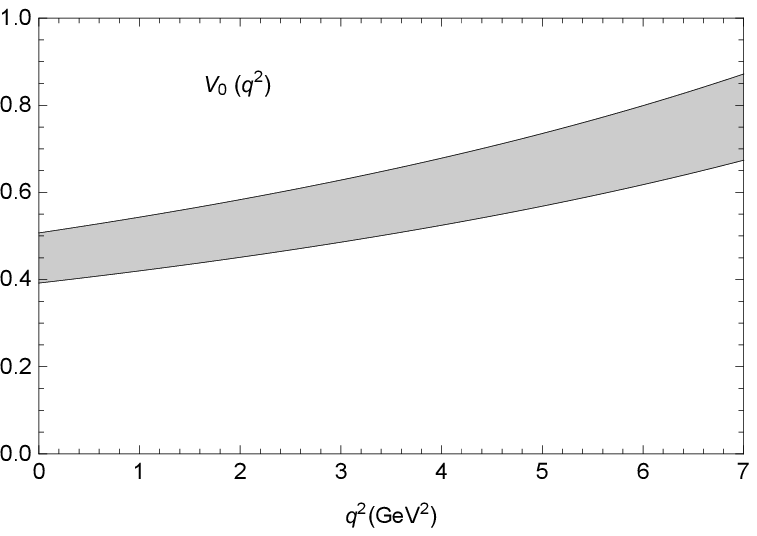}
\includegraphics{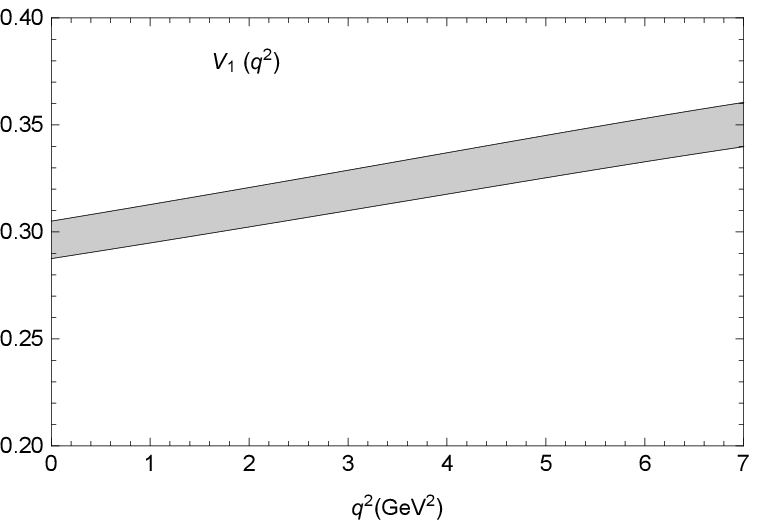}
\includegraphics{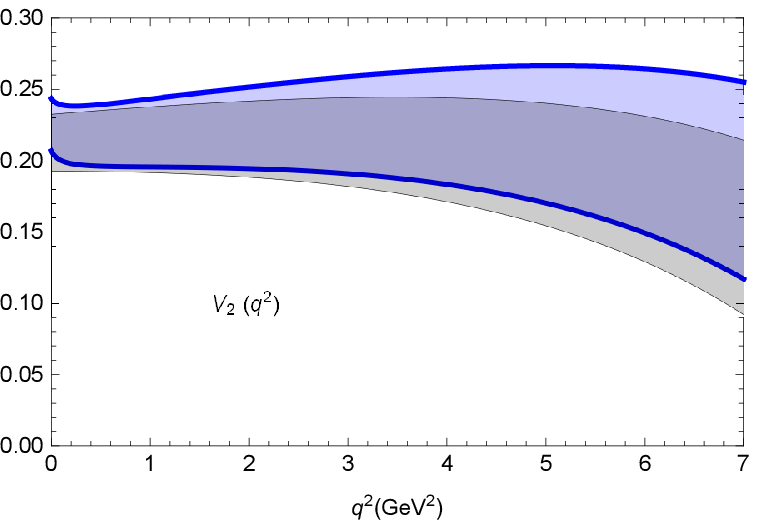}
\includegraphics{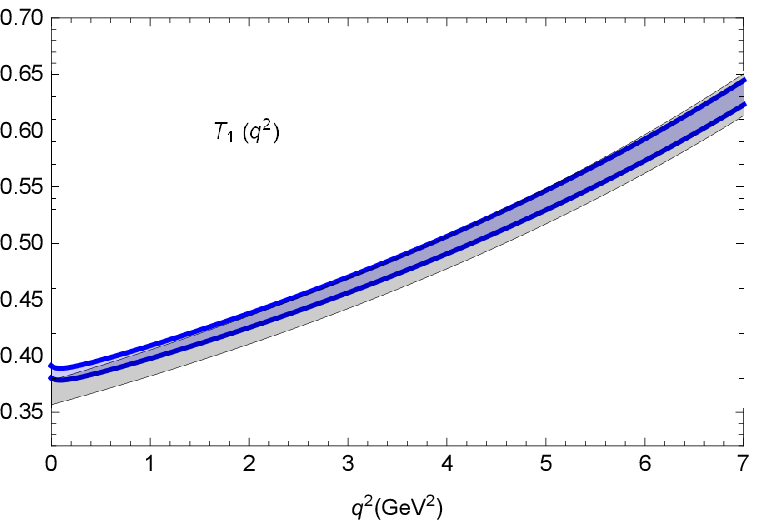}
\includegraphics{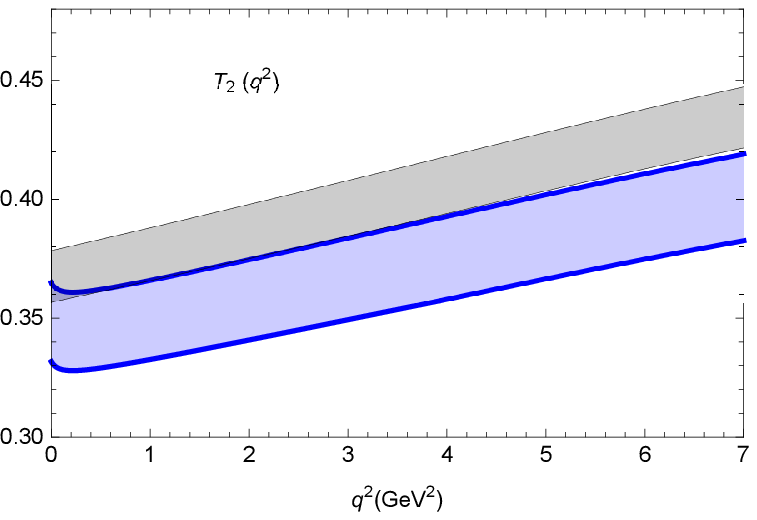}
\includegraphics{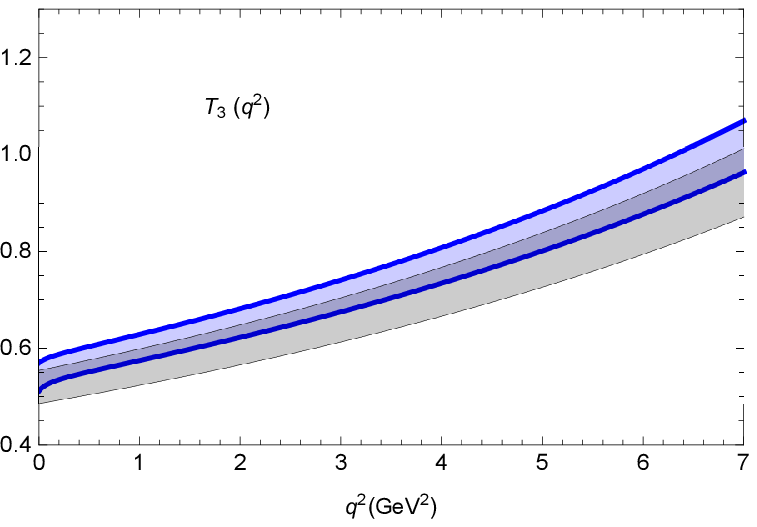}
\caption{Form factors are plotted with $q^2$. Using the uncertainties in the form factors calculated in \cite{Yang:2008zt, Hatanaka:2008ha} at $q^2=0$ along with the other input parameters and by parameterizing them with $q^2$ through Eq. (\ref{parametrize}) their trend without symmetry breaking corrections is shown with black band. The one with blue band corresponds to the same uncertainities in form factor but this time including symmetry breaking corrections. The hard corrections are calculated at $\alpha_{s} =0.34$ at scale $\mu =1.47$ GeV. Tensor form factors are renormalized at the $b-$quark mass i.e., $\mu = m_b $.}
\label{fig:FORMS}
\end{figure}

\section{Numerical Analysis and Applications}\label{NumAny}
\subsection{The Form Factors Analysis}\label{forms}
In this section, we perform the numerical analysis of the form
factors. To calculate the $\alpha_s$ corrections to the form factors
we use the $B$-mesons decay constant to be
$f_B = 0.195\pm 0.01$ GeV. It has already been mentioned in Eq. (\ref{mix-states}) that the physical states $K_1(1200,\; 1400)$ are the mixture of flavor states $K_{1( A,\; B)}$, therefore, the decay constant corresponding physical states $f_{K_1}^{\perp} (f_{K_1}^{\parallel})$ can be obtained by mixing $f_{K_{1(A,B)}}^{\perp}= 0.122 ,0.0884 (f_{K_{1(A,B)}}^{\parallel}= 0.17,0.125$) GeV \cite{Yang:2007zt}. All quantities in hard scattering amplitudes are calculated at the
scale of $1.5$ GeV. For the flavor states, we expand $\phi_{K_1}^{\perp,\parallel}$ up to second Gegenbauer moment \cite{Yang:2007zt}:
\begin{eqnarray}
\phi_{K_{1A}} ^{\perp}&=&6u\bar{u}\left(a_{0A} ^{\perp}+3a_{1A}^{\perp}(2u-1)+\frac{3}{2}a_{2A} ^{\perp}\left(5((2u-1)^2-1\right)\right),\nonumber\\
\phi_{K_{1B}} ^{\perp}&=&6u\bar{u}\left(1+3a_{1B}^{\perp}(2u-1)+\frac{3}{2}a_{2B} ^{\perp}\left(5((2u-1)^2-1\right)\right),\nonumber\\
\phi_{K_1A} ^{\parallel}&=&6u\bar{u}\left(1+3a_{1B}^{\parallel}(2u-1)+\frac{3}{2}a_{2B} ^{\parallel}\left(5((2u-1)^2-1\right)\right),\nonumber\\
\phi_{K_1B} ^{\parallel}&=&6u\bar{u}\left(a_{0B}
^{\parallel}+3a_{1B}^{\parallel}(2u-1)+\frac{3}{2}a_{2A}
^{\parallel}\left(5((2u-1)^2-1\right)\right),\label{twist-3}
\end{eqnarray}
and the values of $a^{\perp, \parallel}_{0,1,2}$ are given in \cite{Yang:2007zt}. In case of the soft form factors, we need the numerical values of the
functions $\xi^{\parallel}_{K_1}$, $\xi^{\perp}_{K_1}$. For this, we
parametrize them in terms of the energy in the large
recoil limit as
\begin{equation}
\left(\frac{m_B+m_{K_1}}{m_B}\frac{E_F}{\Delta}\xi^\perp_{K_1}
,\frac{E_F}{m_{K_1}}\xi^\parallel_{K_1}\right) =\left\lbrace
A(0),V_0(0)\right\rbrace\times\left(\frac{m_B}{2E_F}\right)^2 \label{parametrize}
\end{equation}
where we are going to use the LCSR form factor values $\left(A_{K_{1A}}(0)= 0.45\pm0.09 ,A_{K_{1B}}(0)= -0.37^{+0.10}_{-0.06}\right)$ and
$\left(V_{0(K_{1A})}(0)=0.22\pm 0.04,V_{0(K_{1B})}(0)=-0.45^{+0.12}_{-0.08}\right)$\cite{Yang:2008zt}. The values of other form factors for these flavor states are calculated in \cite{Yang:2008zt} and are summarized in Table IV of ref. \cite{Hatanaka:2008ha}. It is worth mentioning that the values of the form factors calculated in \cite{Yang:2008zt} are without taking into account
gluon radiative corrections. Also to emphasis here that the form factors of the physical mass states $K_1$(1270,1400) also mix, i.e., 
\begin{eqnarray}
\mathcal{A}^{(1270)}_{i}(0) &=&\mathcal{A}^{K_{1A}}_{i}(0)\sin\theta_{K_1}+\mathcal{A}^{K_{1B}}_{i}(0)\cos\theta_{K_1}\notag\\
\mathcal{A}^{(1400)}_{i}(0) &=&\mathcal{A}^{K_{1A}}_{i}(0)\cos\theta_{K_1}-\mathcal{A}^{K_{1B}}_{i}(0)\sin\theta_{K_1} , \label{form-mix}
\end{eqnarray}
where $\mathcal{A}_i(0)$ can be $A(0),\; V_{0,1,2}(0)$ and $T_{1,2,3}(0)$. The form factors $A_0(q^2)$ and $V_0(q^2)$ correspond to our renomalization convention to denote perpendicular $\left(\xi^\perp_{K_1}\right)$ and parallel $\left(\xi^\parallel_{K_1}\right)$ components, respectively and $V_1(q^2)$ does not receive any radiative corrections. The rest of the form factors $V_{2}(q^2),\; T_{1,2,3}(q^2)$ do have the contributions from the symmetry breaking terms.

Using these numerical values of different input parameters at
$\alpha_s=0.34$ and $\mu=1.47$ GeV, the form factors against
momentum transfer $q^2$ are plotted in Fig. \ref{fig:FORMS}. As an input we used the values of the form factors calculated in \cite{Yang:2008zt} where it can be seen that their values for the states $K_{1A,1B}$ at $q^2=0$ given in Table IV of \cite{Hatanaka:2008ha} are prone by the uncertainties. Using these input values along with the other parameters, the form factor extrapolated with $q^2$ using Eq. (\ref{parametrize}) are plotted in Fig. \ref{fig:FORMS}. In these plots, the black band correspond to the uncertainties in the LCSR form factors without symmetry breaking corrections. The blue band correspond to the same form factors and uncertainties but this time including the symmetry breaking corrections calculated here. It is to be kept in consideration that the tensor form factors are renormalized at $\mu=m_b$. In Fig. \ref{fig:FORMS} we can see that in most of the $q^2$ region the two bands overlaps significantly showing that the symmetry breaking corrections are masked by the uncertainties inherited through the input values of the form factors and other parameters. The most prominent effects at almost all $q^2$ range comes in the tensor form factors $T_2(q^2)$ and $T_3(q^2)$. Quantitatively, we can see that the symmetry breaking corrections change the LEET form factors $V_2(q^2)$ and $T_{1,2,3}(q^2)$ by less 10\%. It is worth mentioning that the major uncertainty lies in hard-spectator corrections due to LCDA of the $B$-meson. In past, due to non-availability of constraints on $\lambda_B$ this uncertainty could rise as high as $\pm$50\% \cite{Beneke:2000wa}. These uncertainties were constrained by BABAR analysis of $B\rightarrow \gamma l\nu$ \cite{Aubert:2009ya} at small recoil. This can further be improved by a similar analysis by BABAR for large recoil radiative decay. The BABAR experiment has  put upper limit of $\lambda_B \sim 669$ MeV and lower limit of $\lambda_B \sim 300$ MeV. Their analysis was further improved in ref. \cite{Beneke:2011nf} as the former does not consider highly energetic photons and radiative/power corrections. For our calculations of form factors for semileptonic decay, the value of $\lambda_B \sim 0.35$ GeV seemed to be optimal. In context of the study \cite{Beneke:2011nf}, we expect more uncertainty at large recoil than at small recoil. That is the reason why radiative corrections become important for precision calculation especially at the regime of $q^2\sim 1 - 3$ GeV$^2$ where the symmetry breaking overlaps less significantly with uncertainties only for the form factors $T_{2,3}(q^2)$. The constraint on $B$-meson light cone distribution amplitude along with uncertainty in $K_1$ decay constant makes our results susceptible to $\pm$25\% uncertainty.

\subsection{Applications}\label{applications}
To see how the symmetry breaking corrections influence the values of
observables, we study the implications of the modified form factors on the zero-position of the forward-backward asymmetry and the longitudinal lepton polarization asymmetry $(P_L)$ for the decay channel $B\rightarrow K_1 \mu^+ \mu^-$.
\subsubsection{Forward-backward Asymmetry}
\begin{figure}
 \includegraphics[width=100mm]{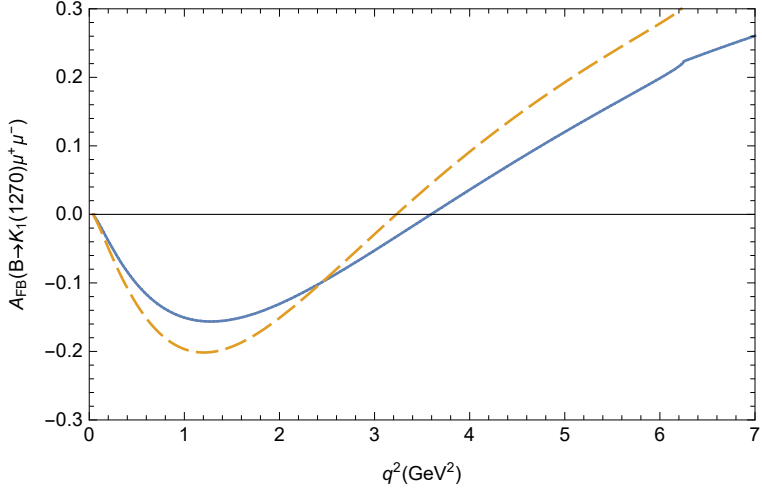}
  \caption{Forward-backward asymmetry as a function of $q^2$ is plotted. The solid and dashed lines correspond to the form factors without and with corrections, respectively. }
  \label{fig:FBA}
\end{figure}

The forward-backward asymmetry and its zero position in $B\rightarrow K_1 \ell^+ \ell^-$ provides an interesting tool to look for the physics beyond the SM. At leading order in the SM, this asymmetry has very weak dependence on the form factors. It is, therefore, interesting to see the effects of form factors incorporating the symmetry breaking corrections on the behavior and the zero-position of the forward-backward asymmetry.  The effective Hamiltonian responsible for the decay under consideration is
\begin{equation}
\qquad\mathcal{H}=\frac{G_F}{\sqrt{2}}V_{ts} ^* V_{tb}\sum_{i=1} ^{10}C_i(\mu)\mathcal{O}_i(\mu),
\end{equation}
where $\mathcal{O}_i$ is are four-quark local operators and $C_i$ are Wilson coefficients calculated in Naive dimensional regularization (NDR) scheme at a scale $\mu$.
At the quark level, the corresponding amplitude for the underlying transition $b\to s \ell^{+}\ell^{-}$ is
\begin{equation}
\mathcal{M}(b\rightarrow s \ell^+ \ell^-)=\frac{G_F \alpha}{\sqrt{2}}V_{ts} ^* V_{tb}\left[C_9 ^{eff}\left(\bar{s}\gamma_{\mu}L b\right)\left(\bar{\ell}\gamma^\mu \ell\right)+C_{10} \left(\bar{s}\gamma_{\mu}L b\right)\left(\bar{\ell}\gamma^\mu \gamma^5 \ell\right)-2\frac{m_b}{q^2}C_7 ^{eff}\left(\bar{s}i\sigma_{\mu\nu}\frac{q^\nu}{q^2} R b\right)\left(\bar{\ell}\gamma^\mu \ell\right)\right] \label{qlevel-amp}
\end{equation}
where  $L=\frac{1-\gamma_5}{2},R=\frac{1+\gamma_5}{2}$, with $m_b$ the mass of $b$-quark and $C_7 ^{eff}=C_7 -C_5 /3 -C_6$. $C_9 ^{eff}$ contains both short distance and long distance contributions, given by
\begin{eqnarray}
C_9 ^{eff}(q^2)= C_9(\mu)+Y_{\text{pert}}(\hat{s})+Y_{\text{LD}}(q^2)
\end{eqnarray}
here $\hat{s}= \frac{q^2}{m_b ^2}$. $Y_{\text{pert}}$ represents the perturbative contributions, and $Y_{\text{LD}}$ is the long-distance part. The $Y_{\text{pert}}$ is given in \cite{Buras:1994dj};
\begin{eqnarray}
\qquad Y_{\text{pert}}=h(\hat{m}_c, \hat{s})C_0 -\frac{1}{2}h(1, \hat{s})(4C_3 + 4C_4 + 3C_5 + C_6)\nonumber\\
 -\frac{1}{2}h(0,\hat{s})(C_3 + 3C_4) 
+\frac{2}{9}(3C_3 + C_4 + 3C_5 + C_6)
\end{eqnarray}

As in the LEET, we are working below the $J/\psi$ resonances i.e, di-lepton invariant mass of up to $q^2 =7$ GeV, therefore, we will ignore the contribution from $Y_{\text{LD}}$ . The study of the branching fraction and asymmetries in the decay under consideration is a bit complicated due to the mixing of $K_{1A}$ and $K_{1B}$ states as already pointed out in Eq. (\ref{mix-states}). 

The amplitude of the decay $B\rightarrow K_1 \mu^+ \mu^-$ is found by sandwiching the different $\left(\bar{s}\Gamma_\mu b\right)$ currents between the $B$ and $K_{1}$-mesons and expressing them in terms of form factors as given in Eq. (\ref{mat-Ele1}). Doing the standard procedure, the corresponding partial decay width can be given as 
\begin{eqnarray}
d\Gamma  (B\rightarrow K_1 \mu^+ \mu^-)= \frac{\sqrt{\lambda}}{1024\pi^4 m_B^3}d\cos\theta dq^2 |\mathcal{M}(B\rightarrow K_1 \mu^+ \mu^-)|^2,
\end{eqnarray}
where $\theta$ is the angle between B meson and $\mu^+$. The quantity $\lambda$ is given as \cite{Paracha:2007yx}
\begin{eqnarray}
\lambda \equiv \lambda(q^2, m_{B}^2, m_{K_1}^2)=\left[\left(1-\frac{q^2}{m_{B}^2}\right)^2-\frac{2m_{K_1}^2}{m_{B}^2}\left(1-\frac{q^2}{m_{B}^2}\right)+\frac{m_{K_1}^4}{m_{B}^4}\right]^{1/2}.
\end{eqnarray}
The differential forward-backward asymmetry is defined as;
\begin{eqnarray}
\frac{d\mathcal{A}_{FB}}{dq^2}=\int_0 ^1 d(cos\theta)\frac{d^2\Gamma}{dq^2 d\cos\theta}-\int_{-1} ^0 d(\cos\theta)\frac{d^2\Gamma}{dq^2 d\cos\theta}.
\end{eqnarray}
 The differential decay width can be calculated by expressing the matrix elements of $\langle K_1|\left(\bar{s}\Gamma_\mu b\right)|B\rangle$ using quark level currents given in Eq. (\ref{qlevel-amp}) in which the hadronic part is parametrized in terms of the form factors. Consequently, the normalized differential forward-backward asymmetry takes the form
\begin{eqnarray}
\qquad\frac{d\mathcal{A}_{FB}}{dq^2}=-\frac{1}{d\Gamma / dq^2}\frac{G_F ^2\mid V_{ts}^* V_{tb}\mid^2}{128\pi^3}m_{B}^3 \lambda (q^2 ,m_{K_1}^2)^2\left(\frac{\alpha}{4\pi}\right)^2\frac{8q^2}{m_{B}^2}C_{10} V_1(q^2)A(q^2)\nonumber\\
\times \Re\left[C_9 ^{eff}+\frac{m_b}{q^2}C_7 ^{eff}\left((m_{B} +m_{K_1})\frac{T_1(q^2)}{A(q^2)}+(m_{B} -m_{K_1})\frac{T_2(q^2)}{V_1(q^2)}\right)\right].\label{FB-asymmetry}
\end{eqnarray}
The expression of $\mathcal{A}_{FB}$  given in Eq. (\ref{FB-asymmetry}) involves the ratio of the form factors $\frac{T_{1}(q^2)}{A(q^2)}$ and $\frac{T_{2}(q^2)}{V_{1}(q^2)}$. From Eqs. (\ref{Aqff} - \ref{T3ff}) we can see that the symmetry breaking corrections appear in the form factors $T_{1}(q^2)$ and $T_{2}(q^2)$.  Hence, we can expect the deviation in the  numerical values of both amplitude and zero-position of $\mathcal{A}_{FB}$ from the HQET form factors and Fig.\ref{fig:FBA} depicts this fact. It can be noticed that the shift of the zero-position of $\mathcal{A}_{FB}$ is about 10\% indicating that before attributing any deviation in the zero-position of the $\mathcal{A}_{FB}$ as a NP, it is important to take into account the shift arises due to the symmetry breaking corrections in the form factors. 
\subsection{Longitudinal Lepton Polarization}
\begin{figure}
  \includegraphics[width=100mm]{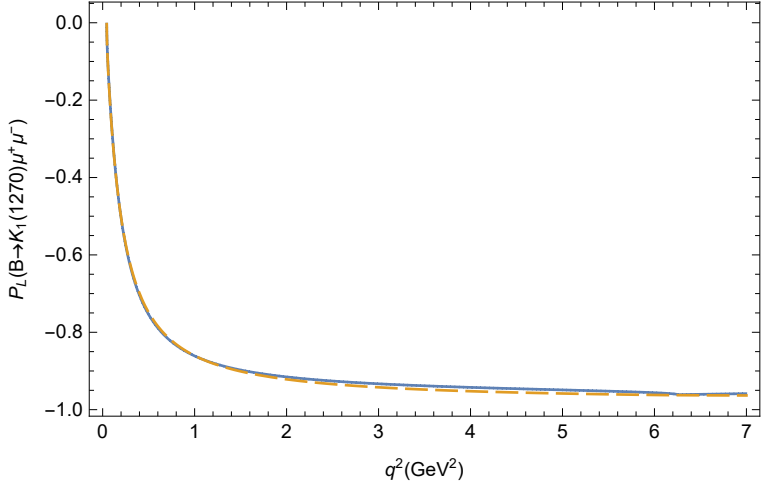}
  \caption{Longitudinal lepton polarization Asymmetry due to symmetry breaking corrections. The description of the solid and dashed lines is the same as in Fig. \ref{fig:FBA}}
  \label{fig:LPAMuons}
\end{figure}
 In principle, many angular observables can be conceived; however, we are interested in longitudinal lepton polarization of the lepton pair. As we get the lepton pair from either off-shell photon, $Z$ boson or some other neutral vector boson; the vertex of the decay to lepton pair has Lorentz structure of either $(V-A)$ or $(V+A)$. Therefore, we can assign different combination of possible helicities and they are summarized in Appendix \ref{AppendixB}. The decay amplitude in term of the lepton and hadron helicity amplitudes can be written as \cite{Li:2009rc}
\begin{eqnarray}
\mathcal{M}(B\rightarrow K_1\mu^+ \mu^-)=-\sum_{i}\left(\mathcal{L}(L,i) \mathcal{H}(L,i)-\mathcal{L}(R,i) \mathcal{H}(R,i)\right),\label{45}
\end{eqnarray} 
where $\mathcal{L}(L)=\bar{\mu}\gamma_\mu(1-\gamma_5)\mu$ and $\mathcal{L}(R)=\bar{\mu}\gamma_\mu(1+\gamma_5)\mu$ are the lepton pair currents. After integrating out $\theta$ and $\phi$ which are defined in the rest frame of lepton pair (c.f. Appendix B) we get the following result
\begin{eqnarray}
\frac{d\Gamma_i}{dq^2}=\frac{\sqrt{\lambda}}{96\pi^3 m_{B}^3}\left[|\mathcal{H}(L,i)|^2+|\mathcal{H}(R,i)|^2\right],
\end{eqnarray}
with $\mathcal{H}(L,i)$, $\mathcal{H}(R,i)$ are the hadronic transition amplitudes and these are summarized in Appendix \ref{AppendixB}. The asymmetry in longitudinal-lepton polarization is written as
\begin{eqnarray}
P_L =\int_0 ^1 d\cos\theta_1 \left(|\mathcal{L}(L,0)\mathcal{H}(L,0)|^2\right) -\int_{-1} ^0 d\cos\theta_1 \left(|\mathcal{L}(R,0)\mathcal{H}(R,0)|^2\right)\label{PL},\label{Long-helicity}
\end{eqnarray}  
where $|\mathcal{L}(R,0)|^2=|\mathcal{L}(R,0)|^2=4q^2 \sin^2 \theta_1$. Integrating $\theta_1$ in Eq. (\ref{Long-helicity}) and normalizing it with full differential decay rate in the denominator, one gets
\begin{eqnarray}
\bar{P}_L =\frac{|\mathcal{H}(L,0)|^2-|\mathcal{H}(R,0)|^2}{|\mathcal{H}(L,i)|^2+|\mathcal{H}(R,i)|^2}.
\end{eqnarray}
Upon subsituting the expressions of $\mathcal{H}(L,i)$ and $\mathcal{H}(R,i)$ the result for the lepton-polarization asymmetry reads as following
\begin{eqnarray}
 P_L &=& -4C^{eff}_9 C_{10}\mathcal{C'} ^2 |V_1(q^2)|^2-4C^{eff}_9 C_{10}\mathcal{D'} ^2 |V_2(q^2)|^2 +4C^{eff}_9 C_{10}\mathcal{C'}\mathcal{D'} V_1(q^2)V_2 (q^2)-
 4C_7C_{10}\mathcal{B'} \mathcal{C'} T_2(q^2)V_1(q^2)\notag\\
&&-4C_7 C_{10} \mathcal{A'}\mathcal{D'} T_3(q^2)V_2(q^2)
 +4C_7 C_{10}\mathcal{B'}\mathcal{D'} T_2(q^2)V_2(q^2)
\end{eqnarray}
where the quantities $\mathcal{A'},\mathcal{B'},\mathcal{C'},\mathcal{D'},$ are given as
\begin{eqnarray}
\mathcal{A'}&=&\frac{\lambda}{m_B^2 -m_{K_1} ^2}, \quad\quad\quad\quad\quad\quad\quad \quad\quad \mathcal{B'}=3m_{K_1} ^2 +m_B^2-q^2 ,\notag \\
\mathcal{C'}&=&(m_B-m_{K_1})(m_{K_1} ^2 -m_B^2 +q^2), \quad\quad \mathcal{D'}=\frac{\lambda}{m_B-m_{K_1}}.
\end{eqnarray} 
Now using the form factors from Eqs. (\ref{Aqff} - \ref{T3ff}) we get $P_L$ in terms of two soft form factors $\xi^{\perp,\parallel}_{K_1}$ and hard spectator factors $\Delta F_{\perp,\parallel}$. The behavior of $P_L$ as a function of di-lepton mass squared is shown in Fig. \ref{fig:LPAMuons}.  As can be observed, there is no difference arise in the value after incorporating the symmetry-breaking corrections to the form factors for this particular observable. Therefore, any significant deviation from the SM prediction of this physical observable in $B\to K_1 \ell^{+}\ell^{-}$ decay will be a hint of a new physics.

\section{Conclusion}\label{conclusion}

In this work, radiative corrections to form factors at one loop order are calculated in $B\to K_1\mu^{+}\mu^{-}$ decay. These corrections are significant at large recoil $q^2 \sim 1-7$ GeV$^2$ for heavy-to-light transitions. We employed a factorization scheme in context of the LEET to take into account the soft- and hard-gluon exchanges. The vertex corrections are found by matching effective theory with full theory at one loop level. These corrections do not break symmetry relations and appear as an $\alpha_s$ corrections in the form factors (c.f. Eq.(\ref{newff})). The hard-spectator corrections do break symmetry relations and these are calculated via light cone distribution amplitudes.  We found that the accumulated corrections to form factor relations shifts the zero-position of the forward-backward asymmetry by 10\%. Therefore, we can say that these symmetry breaking, if not calculated, somehow would have been mixed with the possible NP for this observable in $B\to K_{1}\mu^{+}\mu^{-}$ decay. Contrary to the forward-backward asymmetry, the longitudinal lepton polarization asymmetry hardly gets affected by these symmetry breaking corrections. Therefore, any significant difference especially in the longitudinal lepton polarization asymmetry, if observed experimentally, would be an indicative of some physics beyond the SM.
\section*{Acknowledgments}
 M. J. A would like to thank Prof. Andrzej Czarnecki for giving an opportunity to work as a visiting professor in his group at the University of Alberta during the sabbatical leave from the Quaid-i-Azam University. The work of M. J. A. is partly supported by the Natural Sciences and Engineering Research Council of Canada.
\pagebreak
\appendix
\setcounter{section}{0}
\section{Parton Distribution Amplitudes}\label{AppendixA}
\subsection*{B-meson Parton Distribution Amplitude}
The two particle light cone matrix element with $B$-momentum $m_{B}v$ and two functions $\phi_{\pm}^B (t)$ in coordinate space compatible with Lorentz-decomposition is \cite{Grozin:1996pq}:
\begin{equation}
 M(z)\equiv\langle 0|\bar{q}_\beta (z)P(z,0)q_\alpha (0)|\bar{B}(p)\rangle =-\frac{i f_B m_{B}}{4}\left[\frac{1+\slashed{v}}{2}\left\lbrace2\phi_+ ^B (t)+\frac{\phi_- ^B (t)-\phi_+ ^B (t)}{t}\slashed{z}\right\rbrace\gamma_5\right]_{\alpha\beta}.\label{A1}
\end{equation}
The factor $-\frac{i f_B m_{B}}{4}$ is chosen according to the normalization of pseudo-scalar meson, i.e., $\langle 0|\bar{q}_\beta [\gamma_5]_{\beta\alpha} q_\alpha |\bar{B}(p)\rangle$ and $t=v\cdot z$. The path ordered exponential in Eq. (\ref{A1}) is given as
\begin{equation}
P(z_2,z_1)=P\exp\left(ig_s \int_{z_2} ^{z_1}dz^\mu A_\mu (z)\right).\label{A2}
\end{equation}
Finding the momentum space projector $M^B$ of $M(z)$,
\begin{eqnarray}
\int d^4 z M(z)A(z)&=\int\frac{d^4l}{(2\pi)^4}A(l)\int d^4z e^{-ilz}M(z),\notag\\
&=\int_0 ^\infty dl_+ M^B A(l)|_{l=(l_+ /2)n_+},\label{A3}
\end{eqnarray}
here $A(z)$ is hard scattering amplitude in coordinate space whereas $A(l)$ is its momentum representation. Now, being consistent with our definition of $l$, i.e.,
\begin{equation}
l^\mu =\frac{l_+}{2}n_+ ^\mu +\frac{l_-}{2}n_- ^\mu +l_\perp ^\mu ,\label{A4}
\end{equation}
also the coordinate function $\phi^B _{\pm}(t)$ in momentum space is
\begin{equation}
\phi_{\pm} ^B (t)\equiv \int_0 ^\infty d\omega e^{-i \omega t}\phi_{\pm} ^B (\omega).\label{A5}
\end{equation}
In the heavy quark limit, the hard scattering amplitude $A(l)$ in the light-meson in the $n_-$ direction is independent of $l_-$. So $A(l)=A^0(l_+)+l_\perp ^\mu A^1_\mu(l_+)+O(1/m_B)$. Moreover the derivative is given after dropping the $l_-$ term;
\begin{equation}
\frac{\partial}{\partial l_\mu}=n_-^\mu \frac{\partial}{\partial l_\mu}+ \frac{\partial}{\partial l_{\perp\mu}}. \label{A6}
\end{equation}
Substituting Eqs. (\ref{A5}) and (\ref{A6}) along with $A(l)$ in Eq. (\ref{A3}), we find
\begin{equation}
 M^B _{\beta\alpha}=-\left.\frac{if_B m_B}{4}\left[\frac{1+\slashed{v}}{2}\left\lbrace\phi_+ ^B (\omega)\slashed{n}_+ +\phi_- ^B (\omega)\left(\slashed{n}_- -l_+ \gamma_\perp ^\nu \frac{\partial}{\partial l_\perp ^\nu}\right)\right\rbrace \gamma_5\right]_{\beta\alpha}\right|_{l=(l_+ /2)n_+}. \label{A7}
\end{equation}

\subsection*{$K_1$-meson Parton Distribution Amplitude}

The two parton light-cone distribution amplitude for $K_1$-meson are given as \cite{Yang:2007zt}
\begin{eqnarray}
\langle 0|\bar{q}(y)\gamma_\mu\gamma_5 q(x)|K_1(p',\lambda)\rangle &=& if_{K_1} ^\parallel m_{K_1} \int_0 ^1 du e^{i(up' y+\bar{u}p'x)}\{p'_\mu \frac{\varepsilon^{*(\lambda)}z}{pz}\Phi_{\parallel}(u)+\varepsilon^{*(\lambda)} _{\perp \mu}g_\perp ^{(a)}(u)-\frac{1}{2}z_\mu \frac{\varepsilon^{*(\lambda)}z}{(pz)^2}m_{K_1} ^2 g_3(u)\}\notag\\
\langle 0|\bar{q}(y)\gamma_\mu q(x)|K_1(p',\lambda)\rangle &=& -if_{K_1}^{\parallel} m_{K_1}\epsilon_{\mu\nu\rho\sigma}\varepsilon^{*\nu(\lambda)}p^\rho z^\sigma\int_0 ^1 du e^{i(up'y +\bar{u}p'x)}\frac{g_\perp ^{(v)}(u)}{4}\label{B1}
\end{eqnarray}
here $\Phi_\parallel (u)$ is leading twist-2 distribution amplitudes which can be expanded in Gegenbauer moments as we did in the numerical analysis (c.f. Sec. \ref{NumAny}). $g_\perp ^{(a)}(u),g_\perp ^{(v)}(u)$ are twist-3 while $g_3(u)$ are twist-4 contributions which we did not discuss as there contributions goes as $1/m_{B}^3$. The matrix elements for the tensor currents up to twist-3 are given as
\begin{eqnarray}
\langle 0|\bar{q}(y)\sigma_{\mu\nu}\gamma_5 q(x)|K_1(p',\lambda)\rangle &=& f_{K_1}^{\perp} \int_0 ^1 du e^{i(up'y+\bar{u}p'x)}\{(\varepsilon^{*(\lambda)}_{\perp\mu}p_\nu -\varepsilon^{*(\lambda)}_{\perp\nu}p_\mu)\Phi_{\perp}(u)+ \frac{m_{K_1} ^2\varepsilon^{*(\lambda)}z}{(pz)^2}(p_\mu z_\nu - p_\nu z_\mu)h_\parallel ^{(t)}(u)\},\notag\\
\langle 0|\bar{q}(y)\gamma_5 q(x)|K_1(p',\lambda)\rangle &=& f_{K_1}^\perp m_{K_1} ^2(\varepsilon^{*(\lambda)}z)\int_0 ^1 du e^{i(up'y +\bar{u}p'x)}\frac{h_\parallel ^{(p)}(u)}{2}.
\end{eqnarray}
\setcounter{section}{1}
\section{Helicity Amplitudes}\label{AppendixB}
The polarization vectors are represented by $\varepsilon(i)$ where $i=0,\pm$ denote the longitudinal and transverse polarization of the lepton pair. The metric tensor $g_{\mu\nu}$ can be written in terms of the di-lepton momenta and polarization vectors as; $g_{\mu\nu}=-\sum_i \varepsilon_\mu(i)\varepsilon^* _{\nu}(i)+\frac{q_\mu q_\nu}{q^2}$. 
Substituting it in Eq. (\ref{45}), we can write the decay amplitude as
\begin{equation}
\mathcal{M}(B\rightarrow K_1 \ell^+ \ell^-)=\mathcal{L}_\mu(L)\mathcal{H}_\nu(L)g^{\mu\nu} +\mathcal{L}_\mu(R)\mathcal{H}_\nu(R)g^{\mu\nu} =-\sum_i \mathcal{L}(L,i)\mathcal{H}(L,i)-\sum_i \mathcal{L}(R,i)\mathcal{H}(R,i)\label{B1}
\end{equation}
The leptonic amplitudes are easy to define. Let $\theta_1$ be the angle between $\ell^-$ in the lepton pair rest frame and the $B$-meson. The angle between $K_1$-meson and the lepton pair plane is $\phi$. The various leptonic amplitudes will then be given as
\begin{eqnarray}
\mathcal{L}(L,0)&=2\sqrt{q^2}\sin\theta_1\qquad ,\qquad \mathcal{L}(R,0)=-2\sqrt{q^2}\sin\theta_1\label{B2}\\
\mathcal{L}(L,+)&=-2\sqrt{2}\sqrt{q^2}\sin^2\frac{\theta_1}{2}e^{i\phi}\qquad , \qquad \mathcal{L}(R,+)=-2\sqrt{2}\sqrt{q^2}\cos^2\frac{\theta_1}{2}e^{i\phi}\label{B3}\\
\mathcal{L}(L,-)&=-2\sqrt{2}\sqrt{q^2}\cos^2\frac{\theta_1}{2}e^{-i\phi}\qquad,\qquad \mathcal{L}(R,-)=-2\sqrt{2}\sqrt{q^2}\sin^2\frac{\theta_1}{2}e^{-i\phi}\label{B4}
\end{eqnarray}
The hadronic amplitudes for the three polarization states are given as
\begin{eqnarray}
\mathcal{H}(L,0)=\frac{G_F V_{tb}V^* _{ts}\alpha}{8\sqrt{2}\pi m_{K_1} \sqrt{q^2}}\left(2C_7 ^{eff}m_b \left[\frac{\lambda T_3 (q^2)}{m_B^2 -m^2 _{K_1}}-(3m_{K_1} ^2 +m_B^2-q^2)T_2(q^2)\right]+\right.\nonumber\\\qquad\qquad(C_9 ^{eff} -C_{10}) \left.\left[(m_B-m_{K_1})(m_{K_1} ^2 - m_B ^2+ q^2)V_1(q^2)+\frac{\lambda V_2(q^2)}{m_B-m_{K_1}}\right]\right)\label{B6}\\
\mathcal{H}(R,0)=\frac{G_F V_{tb}V^* _{ts}\alpha}{8\sqrt{2}\pi m_{K_1} \sqrt{q^2}}\left(2C_7 ^{eff}m_b \left[\frac{\lambda T_3 (q^2)}{m_B^2 -m^2 _{K_1}}-(3m_{K_1} ^2 +m_B^2-q^2)T_2(q^2)\right]+\right.\nonumber\\\qquad\qquad(C_9 ^{eff}+C_{10})\left.\left[(m_B-m_{K_1})(m_{K_1} ^2 - m_B ^2+ q^2)V_1(q^2)+\frac{\lambda V_2(q^2)}{m_B-m_{K_1}}\right]\right)\label{B7}\\
\mathcal{H}(L,+)=\frac{G_F V_{tb}V^* _{ts}\alpha}{4\sqrt{2}\pi m_{K_1} \sqrt{q^2}}\left(2 \left[C_7 ^{eff}m_b \sqrt{\lambda} T_1 (q^2)-C_7 ^{eff}m_b (m_B^2 -m_{K_1} ^2)T_2(q^2)\right]+\right.\nonumber\\\qquad\qquad(C_9 ^{eff}-C_{10})q^2\left.\left[\frac{\sqrt{\lambda} A_2(q^2)}{m_B-m_{K_1}}-(m_B-m_{K_1})V_1(q^2)\right]\right)\label{B8}\\
\mathcal{H}(L,-)=\frac{G_F V_{tb}V^* _{ts}\alpha}{4\sqrt{2}\pi m_{K_1} \sqrt{q^2}}\left(2 \left[-C_7 ^{eff}m_b \sqrt{\lambda} T_1 (q^2)-C_7 ^{eff}m_b (m_B^2 -m_{K_1} ^2)T_2(q^2)\right]+\right.\nonumber\\\qquad\qquad(C_9 ^{eff}-C_{10})q^2\left.\left[-\frac{\sqrt{\lambda} A_2(q^2)}{m_B-m_{K_1}}-(m_B-m_{K_1})V_1(q^2)\right]\right)\label{B9}
\end{eqnarray}
\begin{eqnarray}
\mathcal{H}(R,+)=\frac{G_F V_{tb}V^* _{ts}\alpha}{4\sqrt{2}\pi m_{K_1} \sqrt{q^2}}\left(2 \left[C_7 ^{eff}m_b \sqrt{\lambda} T_1 (q^2)-C_7 ^{eff}m_b (m_B^2 -m_{K_1} ^2)T_2(q^2)\right]+\right.\nonumber\\\qquad\qquad(C_9 ^{eff}+C_{10})q^2\left.\left[\frac{\sqrt{\lambda} A_2(q^2)}{m_B-m_{K_1}}-(m_B-m_{K_1})V_1(q^2)\right]\right)\label{B10}\\
\mathcal{H}(R,-)=\frac{G_F V_{tb}V^* _{ts}\alpha}{4\sqrt{2}\pi m_{K_1} \sqrt{q^2}}\left(-2 \left[C_7 ^{eff}m_b \sqrt{\lambda} T_1 (q^2)-C_7 ^{eff}m_b (m_B^2 -m_{K_1} ^2)T_2(q^2)\right]+\right.\nonumber\\\qquad\qquad(C_9 ^{eff}+C_{10})q^2\left.\left[-\frac{\sqrt{\lambda} A_2(q^2)}{m_B-m_{K_1}}-(m_B-m_{K_1})V_1(q^2)\right]\right)\label{B11}
\end{eqnarray}
\setcounter{section}{2}
\section{Hard-Spectator correction to $V_2(q^2)$}\label{AppendixC}
For heavy-to-light meson matrix elements in Eq. (\ref{matrix-elems}), substituting $\mathcal{M}^{K_{1}}$ from Eq. (\ref{K1-projectors}) and $\phi_+ ^B(l_+)$ term of $\mathcal{M}^{B}$ from Eq. (\ref{B-projectors}) along with hard scattering amplitude 
\begin{equation}
\mathcal{T}_{ijkl}^{\mu} = -\left[\gamma^\mu\frac{\slashed{n}_-}{4\bar{u}l_+ m_b E_F}\gamma_\eta\right]_{ij}[\gamma^\eta]_{kl},
\end{equation} 
we arrive at the following relation
\begin{equation}
\left\langle K_1(p',\varepsilon^{*})|\bar{q}\gamma^{\mu}b|\bar{B}(p)\right\rangle_{HSA} = -\frac{4\pi\alpha C_F}{N_c}\left(\frac{-i}{4}\right)\left(\frac{-if_B m_{B}}{8}\right)\left(\frac{1}{4m_b E_F}\right)\int_0 ^1 du \int_0 ^\infty dl_+ \frac{1}{\bar{u}l_+}*\text{Tr}[\cdot\cdot\cdot] \label{C1}
\end{equation}
where the trace is
\begin{equation}
\text{Tr}[\cdot\cdot\cdot]=Tr\left[\left(f^{\perp}_{K_1}\phi_\perp ^{K_1} (u)\slashed{\varepsilon}^*\slashed{p'}+f^{\parallel}_{K_1}\phi_{\parallel}^{K_1} (u)\frac{m_{K_1}}{E_F}(\varepsilon^*\cdot v\slashed{p'})\right)\left(-\gamma^\mu\slashed{n}_-\gamma_\eta\right)\left(\phi_+ ^B(l_+)(1+\slashed{v})\slashed{n}_+\gamma_5\gamma^\eta\right)\right]\label{C2}
\end{equation}
Solving the trace to get
\begin{equation}
\text{Tr}[\cdot\cdot\cdot]=-\frac{8m_{K_1}^2}{E_F}f_{K_1}^\perp \phi_\perp ^{K_1}(u)\phi_+ ^B(l_+)(\varepsilon^*\cdot v)n_-^\mu -32f^{\parallel}_{K_1} \phi_\perp(u)\phi_+ ^B(l_+)(\varepsilon^*\cdot v)\frac{m_{K_1}\Delta}{E_F}n_- ^\mu. \label{trace}
\end{equation}
The leading twist moments $\phi_{\perp,\parallel}^{K_1}(u)$ and $\phi_+ ^B (l_+)$ are integrated and given as in Eq. (\ref{phib}) and Eq. (\ref{phil}), respectively.  Substituting Eq. (\ref{trace}) in Eq. (\ref{C1}) and comparing it to first line in Eq. (\ref{mat-Ele1}) will give us the desired result of Eq. (\ref{V2ff}). The first contribution goes like $m_{K_1}^2$ which will be multiplied by $q^2$ upon comparison with Eq. (\ref{mat-Ele1}) and hence can be omitted. Similar technique can be followed for the calculation of the hard-spectator corrections to the rest of the form factors.


\begin{thebibliography}{99}

\bibitem{Albrecht} J. Albrecht, F. Bernlochner, M. Kenzie, S. Reichert, D. Straub and A. Tully, arXiv: 1709.10308.

\bibitem{Dyk} J. Albrecht, S. Reichert and D. van Dyk, Int. J. Mod. Phys. A {\bf 33},1830016  (2018) [arXiv: 1806.05010].

\bibitem{Dyk2} N. Gubernari, A. Kokulu, D. van Dyk, arXiv: 1811.00983.

\bibitem{Isgur:1989vq}
N.~Isgur and M.~B.~Wise,
Phys.\ Lett.\  B {\bf 232}, 113 (1989). 

\bibitem{Isgur:1989ed}
 N.~Isgur and M.~B.~Wise,
 Phys.\ Lett.\  B {\bf 237}, 527 (1990).
	
	
\bibitem{Neubert:1993mb}
M.~Neubert,
 Phys.\ Rept.\  {\bf 245}, 259 (1994).

\bibitem{Charles:1998dr}
  J.~Charles, A.~Le Yaouanc, L.~Oliver, O.~Pene and J.~C.~Raynal,
  Phys.\ Rev.\  D {\bf 60}, 014001 (1999).
	
\bibitem{Grozin:1996pq}
  A.~G.~Grozin and M.~Neubert,
  Phys.\ Rev.\  D {\bf 55}, 272 (1997).

\bibitem{Georgi:1990um}
  H.~Georgi,  Phys.\ Lett.\  B {\bf 240}, 447 (1990).
  
 \bibitem{Bauer:2002aj} 
  C.~W.~Bauer, D.~Pirjol and I.~W.~Stewart,
  Phys.\ Rev.\ D {\bf 67}, 071502 (2003)
  [hep-ph/0211069].
 
\bibitem{Beneke:2000wa} 
  M.~Beneke and T.~Feldmann,
  Nucl.\ Phys.\ B {\bf 592}, 3 (2001)
  [hep-ph/0008255].
  
  \bibitem{Beneke:2001at} 
  M.~Beneke, T.~Feldmann and D.~Seidel,
  Nucl.\ Phys.\ B {\bf 612}, 25 (2001)
  [hep-ph/0106067].
  
\bibitem{Hatanaka:2008} H. Hatanaka and K. C. Yang, Phys. Rev. D {\bf 77}, 094023 (2008) Erratum: [Phys. Rev. D {\bf 78}, 059902 (2008)].

\bibitem{Paracha:2007yx} 
  M.~A.~Paracha, I.~Ahmed and M.~J.~Aslam,
  Eur.\ Phys.\ J.\ C {\bf 52}, 967 (2007)
  [arXiv:0707.0733 [hep-ph]].

\bibitem{Li:2011nf} 
  Y.~Li, J.~Hua and K.~C.~Yang,
  Eur.\ Phys.\ J.\ C {\bf 71}, 1775 (2011)
  [arXiv:1107.0630 [hep-ph]].
  
  \bibitem{Ju:2014oha} 
  W.~L.~Ju, G.~L.~Wang, H.~F.~Fu, Z.~H.~Wang and Y.~Li,
  JHEP {\bf 1509}, 171 (2015)
  [arXiv:1407.7968 [hep-ph]].
  
  \bibitem{Falahati:2014yba} 
  F.~Falahati and A.~Zahedidareshouri,
  Phys.\ Rev.\ D {\bf 90}, no. 7, 075002 (2014).
  
  \bibitem{Momeni:2018udf} 
  S.~Momeni and R.~Khosravi,
  Eur.\ Phys.\ J.\ C {\bf 78}, no. 10, 805 (2018)
  [arXiv:1805.07046 [hep-ph]].
  
  \bibitem{Momeni:2018tjf} 
  S.~Momeni and R.~Khosravi,
  Phys.\ Rev.\ D {\bf 96}, no. 1, 016018 (2017)
  [arXiv:1804.04844 [hep-ph]].

 \bibitem{Huang:2018}
 Z. -R Huang, M. Ali Paracha, Ishtiaq Ahmed and Cai-Dian Lu
arXiv: 1812.03491 [hep-ph]. 

\bibitem{Yang:2008zt} 
  K.~C.~Yang,
 Phys. Rev. D {\bf 78}, 034018 (2008).
    
\bibitem{Hatanaka:2008ha}  
 H. Hatanaka and K. C. Yang, 
 Phys.\ Rev.\ D {\bf 78}, 074007 (2008).

\bibitem{Ebert:2001pc} 
  D.~Ebert, R.~N.~Faustov and V.~O.~Galkin,
  Phys.\ Rev.\ D {\bf 64}, 094022 (2001)
  [hep-ph/0107065].
  
\bibitem{AJSI}
 Arslan Sikandar, M. Jamil Aslam, Ishtiaq Ahmed and Saba Shafaq, In progress.
 
  \bibitem{Bauer:2000yr} 
  C.~W.~Bauer, S.~Fleming, D.~Pirjol and I.~W.~Stewart,
  Phys.\ Rev.\ D {\bf 63}, 114020 (2001)
  doi:10.1103/PhysRevD.63.114020
  [hep-ph/0011336].
     
\bibitem{Buras:1994dj} 
  A.~J.~Buras and M.~Munz,
  Phys.\ Rev.\ D {\bf 52}, 186 (1995)
  [hep-ph/9501281].

\bibitem{Yang:2007zt} 
  K.~C.~Yang,
  Nucl.\ Phys.\ B {\bf 776}, 187 (2007)
  [arXiv:0705.0692 [hep-ph]].
     
\bibitem{Aubert:2009ya} 
  B.~Aubert {\it et al.} [BaBar Collaboration],
  Phys.\ Rev.\ D {\bf 80}, 111105 (2009)
  [arXiv:0907.1681 [hep-ex]].
  
\bibitem{Beneke:2011nf} 
  M.~Beneke and J.~Rohrwild,
  Eur.\ Phys.\ J.\ C {\bf 71}, 1818 (2011)
  [arXiv:1110.3228 [hep-ph]].
  
\bibitem{Li:2009rc} 
  R.~H.~Li, C.~D.~Lu and W.~Wang,
  Phys.\ Rev.\ D {\bf 79}, 094024 (2009)
  [arXiv:0902.3291 [hep-ph]].   

    
\end{thebibliography}
\end{document}